\documentclass[12pt,reqno]{article}
\usepackage{amsmath}
\usepackage{amsthm}
\usepackage{amsfonts}
\usepackage{amssymb}
\usepackage{epsfig}
\usepackage{graphics}
\input{xy}
\xyoption{all}
\numberwithin{equation}{section}

\def \Fig#1#2#3 {
\begin{figure}
\centering
\epsfxsize=#2cm \epsfbox{#1.eps}
\caption{#3}
\label{#1}
\end{figure}
}

\newcount\figno
\def\fig#1#2#3{
\par\begingroup\parindent=0pt\leftskip=1cm\rightskip=1cm\parindent=0pt
\baselineskip=15pt
\global\advance\figno by 1
\epsfxsize=#3
\centerline{\epsfbox{#2}}
\vskip 12pt
{\bf \small Figure \the\figno:} {\small #1}\par
\endgroup\par
}
\def\figlabel#1{\xdef#1{\the\figno
\mbox{ }}}
\def\encadremath#1{\vbox{\hrule\hbox{\vrule\kern8pt\vbox{\kern8pt
\hbox{$\displaystyle #1$}\kern8pt}
\kern8pt\vrule}\hrule}}

\newcommand{\oh}{{\scriptstyle\frac{1}{2}}}
\newcommand{\gl}{{\mathfrak{gl}}}

\def\e{\epsilon}

\newcommand{\beq}{\begin{equation}}
\newcommand{\eeq}{\end{equation}}

\def\asl{\widehat{sl}}        

\def\sfrac12{{\scriptstyle \frac12}}
\def\P{{\cal P}}

\def\SL2R{{SL(2,$\mathbb{R}$)}}
\def\SL2C{{SL(2,$\mathbb{C}$)}}
\def\SU2{{SU(2)}}

\def\nn{\nonumber}
\def\beas{\begin{eqnarray*}}
\def\eeas{\end{eqnarray*}}
\def\bea{\begin{eqnarray}}
\def\eea{\end{eqnarray}}

\newcommand{\remlst}{\begin{list}
{(\arabic{num})}{\usecounter{num}\topsep0cm \itemsep0cm \parsep0cm}}

\newcommand{\eq}{\begin{eqnarray*}}
\newcommand{\qe}{\end{eqnarray*}}
\newcommand{\eqn}{\begin{eqnarray}}
\newcommand{\qen}{\end{eqnarray}}

\newcommand{\Complex}{\mathbb{C}}

\newcommand{\mat}{\begin{pmatrix}}
\newcommand{\tam}{\end{pmatrix}}

\newcommand{\mc}{\mathcal}
\newcommand{\mf}{\mathfrak}

\def\g{\mathfrak{g}}

\def\P{{\cal P}}

\def\SL21{\text{SU(2$|$1)}}
\def\asl21{\widehat{\text{sl}}\text{(2$|$1)}}
\def\mSL21{\mbox{\rm {SL(2$|$1)}}}
\def\SU2{{SU(2)}}

\def\sl21{\text{sl(2$|$1)}}
\def\gl11{\text{gl(1$|$1)}}
\def\GL11{\text{GL(1$|$1)}}
\def\Mu11{{\mbox{\u11}}}
\def\mGL11{{\mbox{\rm GL(1$|$1)}}}

\title{\bf On the SU(2$|$1) WZNW model \\[3mm]
and its statistical mechanics applications}
\author{Hubert Saleur$^{*}$ and  Volker Schomerus$^{**}$\\[5mm]
$^{*}$Service de Physique Th\'eorique, CEA Saclay,\\
F-91191 Gif-sur-Yvette, France\\[5mm]
$^{**}$DESY Theory Group, DESY Hamburg  \\
Notkestrasse 85, D-22603 Hamburg, Germany\\[5mm]
$^{*}$Physics Department\\
University of Southern California\\
Los Angeles CA 90089-0484, USA\\[5mm]}
\date{Nov. 2006}  

\begin{document}
\begin{titlepage}      \maketitle       \thispagestyle{empty}

\vskip1cm
\begin{abstract}
Motivated by a careful analysis of the Laplacian on the supergroup
$\SL21$ we formulate a proposal for the state space of the $\SL21$
WZNW model. We then use properties of $\asl21$ characters to
compute the partition function of the theory. In the special case
of level $k=1$ the latter is found to agree with the properly
regularized partition function for the continuum limit of the
integrable $\sl21\  3-\bar3$ super-spin chain. Some general
conclusions applicable to other WZNW models (in particular the case 
$k=-1/2$) are also drawn.
\end{abstract}

\vspace*{-19.9cm}\noindent 
{\tt {SPhT-T06/143}}\\
{\tt {DESY 06-201}}
\bigskip\vfill
\noindent
\phantom{wwwx}{\small e-mail:}{\small\tt
hubert.saleur@cea.fr, volker.schomerus@desy.de}
\end{titlepage}

\baselineskip=19pt
\setcounter{equation}{0}
\section{Introduction}

The $\SL21$ WZNW model is a key example of the sigma models with
supergroup targets that appear in the supersymmetric description
of non interacting disordered systems  in low dimensional 
statistical mechanics. The first occurrence of this model probably 
arose via a supersymmetrization of the path integral for two copies 
of the two dimensional  critical Ising model. It was shown in  
\cite{Bernard:1995as} how a $\beta\gamma$ system (with central 
charge $c=-1$) could be introduced to cancel out the pair of  
free Majorana fermions (regrouped for convenience into a Dirac 
fermion) path integrals
\begin{equation}
Z\ =\ \int [d\psi d\psi^{\dagger}d\beta d\gamma]\,  
   \exp[S_{0}+\delta S]\ =\ 1 \label{ising}
\end{equation}
where
\begin{equation}
S_{0}\ =\ \int \frac{d^{2}x}{2\pi} \left(\psi^{\dagger}\bar{\partial}\psi
+\bar{\psi}^{\dagger}\partial\bar{\psi}
+\beta\bar{\partial}\gamma+\bar{\beta}\partial\bar{\gamma}\right)\label{isingfree}
\end{equation}
and
\begin{equation}
    \delta S\ =\ \int \frac{d^{2}x}{2\pi}~~ i\frac{m(x)}{
2}\left(\bar{\psi}^{\dagger}\psi-\psi^{\dagger}\bar{\psi}+
 \bar{\beta}\gamma-\beta\bar{\gamma}\right)\ \ . 
\end{equation}
The theory without random mass $m(x)=0$ is obviously a free
OSP(2$|$2) theory, which can be considered as a $\SL21$ WZNW model at
level $k=-1/2$ \footnote{Our conventions are such that the
sub SU(2) algebra has level $k$. In part of the literature, the level
is defined as $-2\times \hbox{ ours}$, so the free system in
(\ref{ising}) has $k=1$ there.}. Averaging over disorder produces a
marginally irrelevant
current current perturbation of this WZNW model. This is crucial to
understanding the (logarithmic) corrections to pure Ising model scaling.
The deep infrared (IR) behavior however is not changed by the disorder, 
which corresponds to the fact that (\ref{isingfree}) is a simple free 
theory, with pure fermionic correlators identical to those of the usual 
Ising model.
\smallskip 

The second occurrence of the $\SL21$ model is more involved. It
arises in the study of $2+1$ dimensional spin-full electrons in the
presence of a random (non abelian) gauge potential. The
supersymmetrization of the path integral for two copies of the Dirac
fermions produces a free OSP(4$|$4) theory which has been argued to
flow to the $\SL21$ WZNW model at level one under the action of the 
disorder \cite{Bhaseen:2000mi}. The nature of the spectrum and
correlation functions play an important role in the description of
the electronic wave functions at that fixed point. 
\medskip 

Previous works on the $\SL21$ model have focused on some correlation
functions \cite{Maassarani:1996jn},\cite{Ludwig:2000em} and on the 
construction of some characters \cite{Bowcock:1997ce}, but a complete picture of the theory 
has been missing.
\smallskip 

Indeed, the analysis of WZNW on supergroups is notoriously difficult,
even for the simplest case of $\GL11$ \cite{Rozansky:1992rx}. In a recent
paper \cite{Schomerus:2005bf}, we have shown how a careful study of the 
particle limit (in particular, of the simultaneous left and right 
invariant actions
on the space of functions on the group) could provide considerable
insight into this problem. Combining this insight  with some additional
input from the representation theory of the  current
algebras allowed us to  formulate a complete  proposal for the state
space of the theory in the case of  $\GL11$.
The latter involves a rather intricate mixing
of left and right movers that is intimately related to the
representation theory of Lie superalgebras, in particular to the
importance of indecomposable representations. We were then
able to check this proposal through an exact
construction of the theory in the continuum formulation.
\smallskip 

The aim of this work is to extend the lessons we have learned in
\cite{Schomerus:2005bf} to a non-abelian setup, using $\SL21$ as
the simplest non-trivial example.\footnote{To be more precise, we
shall consider the universal cover of $\SL21$ in which the
abelian, time-like circle is replaced by the real line. We shall
comment on this in much more detail in section 4.} Once more, the
analysis of the particle limit (section 2) along with some input
from the representation theory of the $\sl21$ current algebra
(section 3) shall provide all the necessary ingredients for the
construction of the field theory state space (section 4), in close
analogy to our previous investigation of the $\GL11$ model. In the
present case we shall not attempt to verify the structure of the
state space through calculations of correlators, though this would
be possible as well (see \cite{Gotz:2006qp}). Instead we shall use
results on an integrable $\sl21$ spin chain to test our continuum
constructions. Such a spin chain was first investigated in 
\cite{Essler:2005ag}
as a discrete version of the $\SL21$ WZNW model. We shall see that
both approaches are consistent. The comparison, however, is a bit
subtle, mainly due to the fact that the supergroup $\SL21$ has an
indefinite metric. While this poses no problem for the (algebraic)
conformal field theory analysis, the computation of the partition
function on the lattice suffers from divergencies which need to be
regularized. We shall do this through some appropriate analytic
continuation. In this sense, our analysis also supports a
particular prescription for extracting information from spin
chains with an indefinite metric.

\section{The minisuperspace analysis}

The aim of this section is to decompose the space of functions on the
supergroup $\SL21$ into (generalized) eigenfunctions of the quadratic
Casimir element in the regular representations. Since the Casimir
commutes with the generator, the eigenspaces may be decomposed into
representation of the Lie superalgebra $\sl21$. It is therefore useful
to have some background on the representation theory of $\sl21$. We
shall review a few known facts below before addressing the harmonic
analysis. More details can be found e.g.\ in \cite{Frappat:1996pb,
Gotz:2005jz}.

\subsection{The Lie superalgebra $\sl21$}

In this subsection we provide a short overview on finite dimensional
representations of $\sl21$. Rather than reproducing a complete list of
such representations we shall focus on those that are relevant below,
namely on Kac modules and the projective covers of atypicals.

\subsubsection{The defining relations of $\sl21$}

The even part $\mf{g}^{(0)} =$  gl(1) $\oplus$ sl(2) of the Lie
superalgebra $\mf{g} =$ $\sl21$ is generated by four bosonic
elements $H$, $E^\pm$ and $B$ which obey the commutation relations
\begin{align} \label{sl211}
  [H,E^\pm]&\ =\ \pm E^\pm\quad,&
  [E^+,E^-]&\ =\ 2H\quad,&
  [B,E^\pm]&\ =\ [B,H]\ =\ 0\ \ .
\end{align}
In addition, there exist two fermionic multiplets $(F^+,F^-)$ and
$(\bar{F}^+,\bar{F}^-)$ which generate the odd part $\mf{g}^{(1)}$.
They transform as $(\pm \oh,\oh)$ with respect to the even subalgebra,
i.e.\
\begin{align} \label{sl212}
  [H,F^\pm]&\ =\ \pm\frac{1}{2}F^\pm&
  [H,\bar{F}^\pm]&\ =\ \pm\frac{1}{2}\bar{F}^\pm\nn\\[2mm]
  [E^\pm,F^\pm]&\ =\ [E^\pm,\bar{F}^\pm]\ =\ 0&
  [E^\pm,F^\mp]&\ =\ -F^\pm&
  [E^\pm,\bar{F}^\mp]&\ =\ \bar{F}^\pm\\[2mm]
  [B,F^\pm]&\ =\ \frac{1}{2}F^\pm&
  [B,\bar{F}^\pm]&\ =\ -\frac{1}{2}\bar{F}^\pm\ \ .\nn
\end{align}
Finally, the fermionic elements possess the following simple
anti-commutation relations
\begin{align} \label{sl213}
  \{F^\pm,F^\mp\}&\ =\ \{\bar{F}^\pm,\bar{F}^\mp\}\ =\ 0&
  \{F^\pm,\bar{F}^\pm\}&\ =\ E^\pm&
  \{F^\pm,\bar{F}^\mp\}&\ =\ B\mp H
\end{align}
among each other. Formulas \eqref{sl211} to \eqref{sl213} provide
a complete list of relations in the Lie superalgebra $\sl21$.

\subsubsection{Kac modules and irreducible representations}

Kac modules \cite{Kac:1977em} are the basic tool in the construction
of irreducible representations. In the case of $\g=$ $\sl21$,
these form a 2-parameter family $\{b,j\}$ of $8j$-dimensional
representations. We may induce them from the $2j$-dimensional
representations $(b-\oh,j-\oh)$ of the bosonic subalgebra $\g^{(0)}$
by applying the pair $F^\pm$ of fermionic elements.
Our label $b\in\Complex$ denotes a gl(1)-charge and spins of
sl(2) are labeled by $j=\oh,1,\dots$. The dual construction
which promotes the fermionic generators $\bar{F}^\pm$ to creation
operators, yields anti-Kac modules $\{\overline{b,j}\}$ ($b$ and
$j$ take the same values as above). The bosonic content of
(anti-)Kac modules may be read off rather easily from their
construction,
\begin{equation}
  \label{typdec}
  \{b,j\}\bigr|_{\mf{g}^{(0)}}
  \ \cong \  \{\overline{b,j}\}\bigr|_{\mf{g}^{(0)}}
  \ \cong \
   (b-\oh,j-\oh)\ \oplus \  (b,j) \ \oplus \ (b,j-1)
  \ \oplus\ (b+\oh,j-\oh)\ . 
\end{equation}
For generic values of $b$ and $j$, the modules $\{b,j\}$ and
$\{\overline{b,j}\}$ are irreducible and isomorphic. At the points
$\pm b=j$, however, they degenerate, i.e.\ the representations are
indecomposable and no longer isomorphic. In fact, Kac and anti-Kac
modules are then easily seen to possess different invariant
subspaces. To be more precise the (anti-)Kac modules $\{\pm j,j\}$
and $\{\overline{\pm j,j}\}$ are built from two atypical
representations such that
\begin{equation}
  \begin{split} \label{Kmdec}
    \{\pm j,j\}:&\qquad\{j\}_\pm\ \longrightarrow\ \{j-\oh\}_\pm\\[2mm]
    \{\overline{\pm j,j}\}:&\qquad\{j-\oh\}_\pm\ \longrightarrow\
     \{j\}_\pm\ \ .
  \end{split}
\end{equation}
The atypical irreducible representations $\{j\}_\pm$ that appear in
these small diagrams are $4j+1$ dimensional. With respect to the
even subalgebra they decompose according to
\begin{equation} \label{atypdec}
  \{j\}_\pm|_{\g^{(0)}} \ =\ \begin{cases}
                  (j,j)\oplus(j+\oh,j-\oh)&,\text{ for }+
             \text{ and }j=\oh,1,\dots\\[2mm]
                  (-j,j)\oplus\bigl(-(j+\oh),j-\oh\bigr)&,
            \text{ for }-\text{ and }j=\oh,1,\dots\ \
                \end{cases}
\end{equation}
For $j=0$, only the trivial representation $(0)$ occurs. It is
also useful to introduce the characters of these representations.
By definition, these are obtained as
$$ \chi_{\cal R}(z,\xi) \ = \ \text{str}_{\cal R} \left( \xi^{B} 
z ^H\right)$$
where the super-trace extends over all states in the representation
${\cal R}$ of $\sl21$. For Kac modules the character is rather simple. 
In fact, it factorizes
$$ \chi_{\{b,j\}}(\xi,z) \ = \ \xi^{b-1/2}\  \chi_f(\xi,z)\
\sum_{l=-j+1/2}^{l=j-1/2}\,  z^l $$
with a fermionic contribution $\chi_f$ that is independent of the
Kac module under consideration,
$$ \chi_f(\xi,z) \ = \ 1 - \xi^{1/2}\, z^{1/2} - \xi^{1/2}\, z^{-1/2}
 + \xi \ \ . $$
The characters of atypical representations can be obtained easily
form their decomposition formulas (\ref{atypdec}). We would like
to pursue a rather different route here that uses the decomposition
(\ref{Kmdec}) of Kac modules into atypicals. The first formula implies
that
\begin{equation}  \label{chardiff}
 \chi_{\{\pm j,j\}}(\xi,z) \ = \ \chi_{\{j\}_\pm}(\xi,z) -
                                   \chi_{\{j-1/2\}_\pm}(\xi,z)\ \ .
\end{equation}
We can solve these equations for the characters of atypical
representations by the following infinite sums
\begin{equation} \label{atypchar}
 \chi_{\{j-1/2\}_\pm}(\xi,z) \ = \ - \sum_{n=0}^{\infty}
   \chi_{\{\pm j\pm n/2,j+n/2\}}(\xi,z) \ \ .
\end{equation}\vspace*{5mm}
\fig{\label{fig1}A graphical illustration of how
characters for an atypical representation can be obtained as an
infinite sum of characters of Kac modules. Here the one
dimensional atypical identity ${0}$ appears as a sum over
$\{1/2,1/2\}$ (thin lines and dots), $\{1,1\}$ (medium lines and
dots), $\{3/2,3/2\}$ (thick lines and dots) etc.\ All spurious
contributions (that is, the whole tower but the origin) appear
twice, and they disappear by cancellations of bosonic and
fermionic degrees of freedom. The diagram corresponds to the
choice of plus sign in formula (\ref{atypchar}).}
{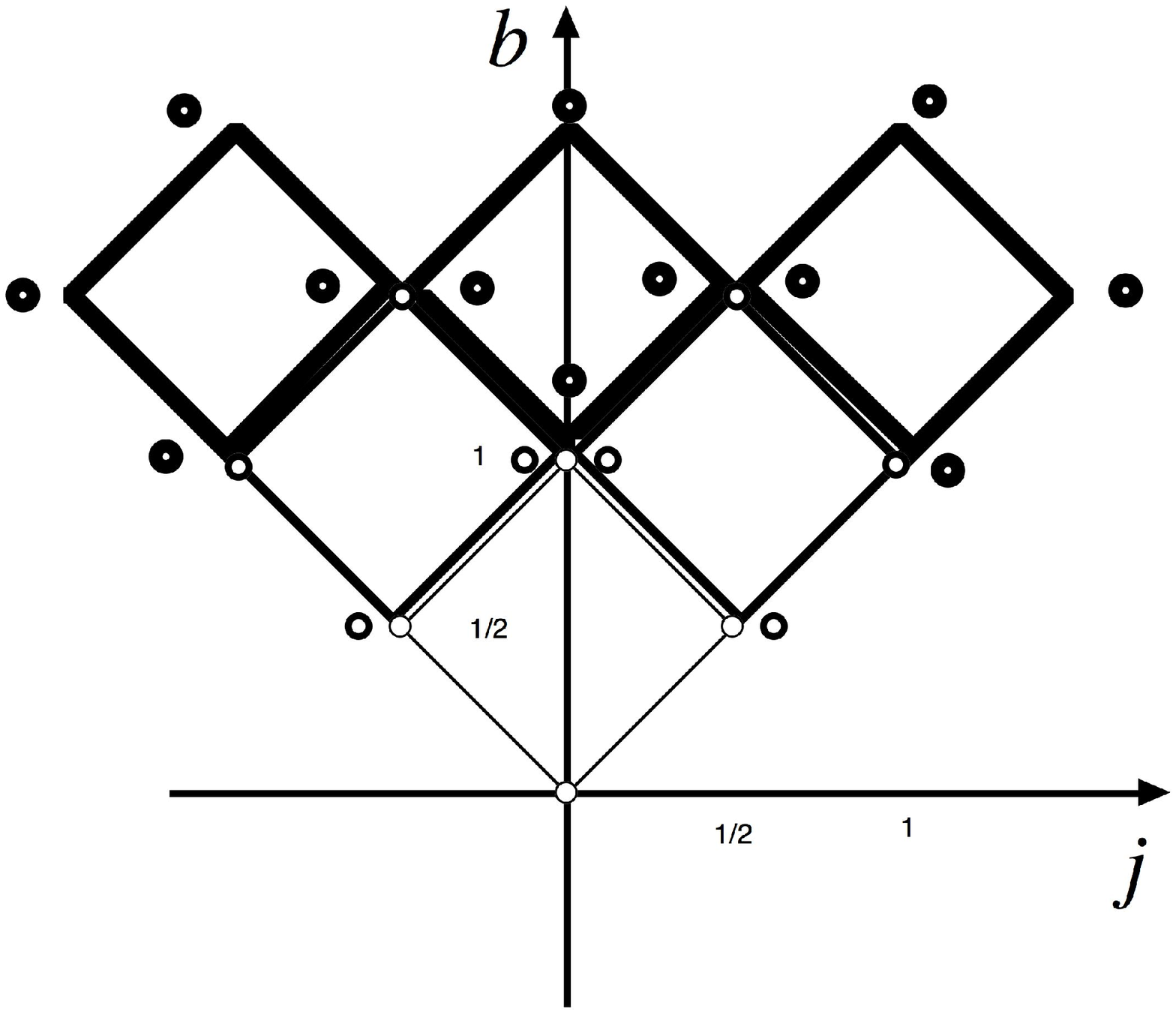}{10truecm} \figlabel{\basic} \vspace{5mm}
One may check by explicit computation that the contributions from
all but two bosonic multiplets cancel each other in the infinite
sum through a mechanism that is visualized in Figure 1. The
remaining two terms certainly agree with the decomposition
formulas (\ref{atypdec}). Our derivation here may seem like a
rather complicated path for such a simple result, but we shall see
later that the same trick works for characters of atypical affine
representations which are otherwise difficult to obtain.

\subsubsection{Projective covers of atypical irreducible modules}

By definition, the projective cover of a representation
$\{j\}_\pm$ is the largest indecomposable representation
$\mc{P}^\pm(j)$ which has $\{j\}_\pm$ as a subrepresentation (its
socle). We do not want to construct these representations
explicitly here. Instead, we shall display how  they are composed
from atypicals. The projective cover of the trivial representation
is an $8$-dimensional module of the form
\begin{equation}
  \mc{P}(0):\qquad\{0\}\ \longrightarrow\ \{\oh\}_+\oplus\{\oh\}_-\
    \longrightarrow\ \{0\}\ \ .
\end{equation}
For the other atypical representations $\{j\}_\pm$ with $j=\oh,1,\dots$
one finds the following diagram,
\begin{equation} \label{P}
  \mc{P}^\pm(j):\qquad\{j\}_\pm\ \longrightarrow\ \{j+\oh\}_\pm
\oplus\{j-\oh\}_\pm\ \longrightarrow\ \{j\}_\pm\ \ .
\end{equation}
These representation spaces are $16j+4$-dimensional. Let us agree
to absorb the superscript $\pm$ on $\mc{P}$ into the argument,
i.e.\ $\mc{P}^\pm(j)= \mc{P}(\pm j)$, wherever this is convenient.

\subsection{Functions on the supergroup $\SL21$}

Now we are prepared to analyze the space of functions on the
supergroup $\SL21$. For this purpose, let us introduce
coordinates through the following explicit decomposition
of elements $U \in \SL21$,
$$ U \ = \ e^{i\bar \eta_\nu \bar F^{\nu}} \, e^{izB} \ g\
   e^{i\eta_\nu F^\nu}
$$
Here, the bosonic base SU(2)$\times \mathbb{R}$ is parametrized
by an element $g \in $ SU(2) $\cong \text{S}^3$ along with the
time-like variable $z$. In these coordinates, the generators
of the right regular action read
\begin{eqnarray}
R_{F^{\pm}} \ = \ -i\partial_\pm  \ \ & , & \ \
R_{E^{\pm}} \ = \ R^0_{E^\pm} + \eta_\mp \partial_\pm
   \\[2mm]
R_H = R_H^0 + \frac12 \eta_- \partial_- - \frac12 \eta_+
         \partial_+\ \ & , & \ \
R_B \ = \ -i\partial_z - \frac12 \eta_-\partial_- -
 \frac12 \eta_+ \partial_+ \ \ , \\[2mm]  & &
\hspace*{-6cm} R_{\bar F^{\pm}} \ = \  i e^{-iz/2} \,
                    D^{1/2}_{\nu (\pm 1/2)} (g) \,
                    \bar\partial_{-\nu}
   + i \eta_\pm (R^0_{E^\pm} + \eta_\mp \partial_\pm)
   - i \eta_\mp ( i \partial_z \mp R^0_H )
\end{eqnarray}
where $R^0_X$ are the generators of the right regular
representation of SU(2). They act on the matrix elements
$D^j_{ab}(g), a,b=-j,-j+1,\dots j,$ according to
\begin{eqnarray*}
 R_H^0 \, D^j_{ab}(g) \ = \ b \, D^j_{ab}(g)\  & , & \
   R_{E^+}^0 \, D^j_{ab}(g) \ = \ \sqrt{(j+b+1)(j-b)}
     \   D^{j}_{a(b+1)}(g) \ \ , \\[2mm]
   R_{E^-}^0 \, D^j_{a(b+1)}(g) & = & \sqrt{(j+b+1)
   (j-b)} \ D^j_{ab}(g) \ \ .
\end{eqnarray*}
Matrix elements with $j=1/2$ appear as coefficients in the
differential operators $R_{\bar F^{\pm}}$ and their behavior
under the action of $R^0_X$ plays an important role in
checking that the above generators of the right regular
representation obey the defining relations of $\sl21$.
Formulas for the left regular representation may be obtained
similarly,
\begin{eqnarray}
L_{\bar F^{\pm}} \ = \ -i\bar \partial_\pm  \ \ & , & \ \
L_{E^{\pm}} \ = \ L^0_{E^\pm} - \bar \eta_\mp \bar \partial_\pm
   \\[2mm]
L_H \ = \ L_H^0 + \frac12 \bar \eta_- \bar \partial_- - \frac12
\bar \eta_+ \bar \partial_+\ \ & , & \ \ L_B \ = \ i\partial_z +
\frac12 \bar \eta_-\bar \partial_- +
 \frac12 \bar \eta_+ \bar \partial_+ \ \ , \\[2mm]  & &
\hspace*{-6cm} L_{F^{\pm}} \ = \  i e^{-iz/2} D^j_{(\pm
1/2)\nu}(g) \ \partial_{-\nu}
   + i \bar \eta_\pm (L^0_{E^\pm} - \bar \eta_\mp \bar \partial_\pm)
   + i \bar \eta_\mp (i\partial_z \mp L^0_H )
\end{eqnarray}
It is probably not necessary to stress that left and right generators
(anti-)commute with respect to each other.
\smallskip

By construction (see however \cite{Gotz:2006qp}), the generators of the
left and right regular representation act on the space of all
Grassmann valued functions with square integrable coefficients
on the bosonic base, i.e.\ on the space
$$ L_2(\SL21) \ := \ L_2(SU(2) \times \mathbb{R}) \otimes
  \Lambda (\eta_\pm,\bar \eta_\pm) $$
where $\Lambda(\eta_\pm,\bar \eta_\pm)$ denotes the Grassmann
algebra that is generated by our four fermionic coordinates
$\eta_\pm$ and $\bar \eta_\pm$. With respect to the left
regular action, the space of square integrable functions
can be shown to decompose as follows,
\begin{eqnarray} \label{Ldec}
L_2(\SL21) &  \cong_L &
 \sum_{j = 1/2}^{\infty}
   \sum_{b \neq \pm j}  4j \left( \{-b,j\} \oplus
    \{b,j\}' \right) \ \oplus \ \\[2mm]
    & &  \hspace*{1cm}  \oplus \ \sum_j \ (2j+1)\,
    \left(\P^+_j \oplus \P^-_j\right) \oplus 2j
    \left(\P^+_j \oplus \P^-_j\right)' \ \ . \nonumber
\end{eqnarray}
Here, the summation runs over $j = 0,1/2,1,\dots$, $\{b,j\}$
denotes the typical representations of $\sl21$ and $\P^\pm_j$ are
the projective covers of the atypical representations $\{j\}_\pm$.
Most of our conventions can be found e.g. in
\cite{Frappat:1996pb}. A prime $'$ on a representation means that
the degree is inverted, i.e.\ that fermionic vectors become
bosonic and vice versa. The result is a special case of the
general observation made in \cite{Huffmann:1994ah} and it
generalizes a similar decomposition we described in
\cite{Schomerus:2005bf} for the left regular action of $\gl11$.
The interested reader can find an explicit  proof in Appendix
A. Let us comment that the decomposition of the left regular
action displays the same violation of the Peter-Weyl theorem as in
the case of $\GL11$. In particular, since the quadratic Casimir is
not diagonalizable in the projective covers ${\cal P}^\pm_j$, the
Laplacian on the supergroup $\SL21$ can only be brought into
Jordan normal form. The blocks can reach a rank up to three.
\smallskip

The functions on our supergroup carry another (anti-)commuting
action of the Lie superalgebra $\g$ by left derivations. There
is a corresponding decomposition which is certainly identical
to the decomposition above. A more interesting problem is to
decompose the space of functions with respect to the graded
product $\g \otimes \g$ in which the first factor acts
through the left regular action while for the second factor
we use the right regular action. In the typical sector, the
$4j|4j$-dimensional multiplicity spaces in the first line of
eq.\ (\ref{Ldec}) get promoted to typical representations of
the right regular action, i.e.\
\begin{equation} \label{LRdec}
L_2(\SL21) \ \cong_{L-R} \
 \sum_{j = 1/2}^{\infty}
   \sum_{b \neq \pm j} \left( \{b,j\}_L \otimes
    \{-b,j\}_R \right) \ \oplus \  {\cal J}
\end{equation}
where ${\cal J}$ is a single indecomposable, containing  all
the atypical building blocks. Its structure may be summarized
by the following picture
{\xymatrixrowsep{40pt}
\xymatrixcolsep{-4pt}
\begin{equation}\label{diagram} 
\xymatrix{ &  \ar[dl] \{\oh\}_-\times \{\oh\}_+ \ar[dr]\ar[drrr] & & & &
             \ar[dl] \ar[dlll] \{0\}\times \{0\} \ar[dr] & \\
 \cdot\cdot \{1\}_- \times \{\oh\}_+ \ar[dr] & & \{0\}\times \{\oh\}_+
 \ar[dl]\ar[drrr] &\ \  &
 \{\oh\}_- \times \{0\} \ar[dr] \ar[dlll] & & \{\oh\}_+\times \{0\}
 \ar[dl] \cdot\cdot \\
  & \{\oh\}_-\times \{\oh\}_+  & & & &
       \{0\}\times \{0\}  &
}
\end{equation}}
This diagram is the natural extension of the corresponding picture
for $\GL11$. It extends to infinity in both directions and combines
all the atypical sectors into a single indecomposable representation.
Note that by construction, each projective cover in the decomposition
of the right regular representation appears with the correct
multiplicity. We shall see below how this picture is modified in
the full quantum theory.

\section{Representation theory of the affine algebra}

The previous analysis of the particle limit applies to all sigma
models on $\SL21$, but the information it provides is usually not
sufficient in order to reconstruct the entire field theory from
it. This is very different for the WZNW model in which the entire
spectrum can be generated from particle wave functions through
current algebra symmetries. We need some facts on the
representation theory of the $\sl21$ current algebra and shall
provide them in the following section. All the results we collect
here are well known from \cite{Bowcock:1996tk,Bowcock:1997ce,
Hayes:1998ku,Semikhatov:2001qx,Semikhatov:2003uc}. Their
derivation, however, is somewhat original. In particular, we shall
use a simple, but highly efficient prescription to construct
characters of atypical representations of $\asl21$ through
infinite sums over typicals. This extends the formula
(\ref{atypchar}) we have discussed in section 2 to an infinite
dimensional setting, thereby generalizing a trick that has first
been proposed in the context of the $\gl11$ current algebra
\cite{Rozansky:1992td}.

\subsection{Some basic ingredients}

Irreducible representations of the affine $\sl21$ algebra can be
built over the irreducible typical representations $\{b,j\}$
with $j=1/2,\dots, k/2$ as well as over the atypicals $\{j\}_\pm$
with $j=0,1/2,1,\dots,k/2$. Ground states in the former set of
representations possess conformal dimension
$$ h_{\{b,j\}} \ = \ (j^2-b^2)/(k+1)\ \  $$
while the conformal dimension for ground states in the latter set
vanishes. Following the work \cite{Bowcock:1996tk} of Bowcock et
al.\ we shall divide these representations into three different
classes. The generic class I representations occur for $\{b, j\}$
with $b \neq j_m^{\pm}$ where we defined
$$ j^\pm_m \ := \ \pm j + m (k+1) \ \ \mbox{ for } \ \ \ m \
 \mbox{ integer} \ \ . $$
Class II representations include those erected over $\{b,j\}$
with $b = j_m^\pm, m \neq 0,$ along with the sectors
generated from atypicals $\{j\}_\pm, j \neq 0$. The vacuum
representation that is generated from the atypical $\{0\}$
is the only member of the final class, which we denote as
class IV for historical reasons.  Our aim is to describe
the singular vectors in the corresponding Verma modules
and to provide the associated formulas for the
super-characters
$$ \chi_{{\cal R}}(q,z
,\xi) \ := \ {\text str}_{{\cal R}}
   \left(\, q^{L_0-\frac{c}{24}}\, \xi^{B_0}\, z^{H_0}\, \right)
$$
of irreducible representations. The results we describe have
first appeared in \cite{Bowcock:1996tk}.
\medskip

Before we start our discussion of characters let us quickly
recall that it is possible to construct $\sl21$ currents in
terms of decoupled bosonic and fermionic variables. To be
more precise, we introduce a set of bosonic currents $e^\pm(z),
h(z),b(z)$ and assume them to satisfy the operator product
expansions of an affine sl(2) algebra at level $k-1$. In
addition, let us introduce two sets of fermionic fields
$p^a$ and $\theta^a$ obeying the canonical relations
$$ \theta^a(z_1)\  p^b(z_2) \ \sim \ \frac{\delta_{ab}}{z_1-z_2}
   + \dots \ \ . $$
Then we can construct an $\sl21$ current algebra at level $k$
through the following prescription,
\begin{equation} \begin{split} \label{BFdec}
E^+(z) & = \ e^+(z) + :\theta^1 p^2:(z) \ \ , \ \ H(z) \ = \
  h(z) + \frac12\, :\left(\theta^1p^1 - \theta^2p^2\right):(z)
\ ,  \ \ \ \ \\[2mm]
E^-(z) & = \ e^-(z) + :\theta^1 p^2:(z) \ \ , \ \ B(z) \ = \
  b(z) - \frac12\, :\left(\theta^1p^1 + \theta^2p^2\right):(z)
\ , \ \ \ \ \\[3mm]
F^+(z) & = \ p^2(z) \ \ \ \ \ , \ \ \ \ \ \ \bar{F}^{+} \ = \
 \theta^2 e^+(z) + \theta^1(b + h)(z)\, -
   :\theta^1 \theta^2 p^2:(z)\ ,   \\[3mm]
  F^-(z) & = \ p^1(z) \ \ \ \ \ , \ \ \ \ \ \ \ \bar{F}^{-} \ = \
 \theta^1 e^-(z) + \theta^2(b - h)(z)\, +
   :\theta^1 \theta^2 p^2:(z)\ . \end{split}
\end{equation}
Since the fermionic fields $\theta^a$ and $p^a$ are supposed to
commute with the bosonic fields $e^\pm(z),h(z)$ and $b(z)$, the
characters of typical representations factorize with the factors
$\theta_1(y,q)$ arising from the fermionic pairs. The shift $j
\rightarrow j-1/2$ in the bosonic contribution may be traced
back to a similar shift in the labeling of typical $\sl21$
representations, see eq.\ (\ref{typdec}).

\subsection{Typical (class I) representations}

The generic class I representations have no singular vectors
except from the ones that arise through the representations of a
bosonic su(2) current algebra at level $k-1$. In this sense, they
may be considered the typical representations of the affine $\sl21$
algebra. The statement implies a precise expression for the
characters of class I representations
\begin{eqnarray} \label{typeI}
\chi^{I}_{\{b,j\}}(q,z,\xi) & = & \frac{ q^{-b^2/(k+1)}}{
\xi^{-b} \eta^{3}(q)} \, \theta_1(z^{1/2}\xi^{1/2},q) \,
  \theta_1(z^{-1/2}\xi^{1/2},q) \
   \chi^{k-1}_{j-1/2} (z,q) \\[2mm]
& & \hspace*{-3cm}  \mbox{ where} \ \ \theta_1(y,q) \ = \
-i y^{1/2} q^{1/8}\  \prod_{n=1}^\infty \,
(1-q^n) (1-yq^n) (1-y^{-1} q^{n-1})
\end{eqnarray}
and $b \neq j^\pm_m$ and $1/2 \leq j \leq k/2$. We also recall
that the su(2) characters are given by
$$ \chi^{k-1}_{j-1/2} (z,q)\ = \
   q^{\frac{j^2}{k+1}-\frac{1}{8}} z^{j-1} \frac{
   \sum_a \, q^{(k+1)a^2+2aj} \left(z^{a(k+1)}-z^{-a(k+1)-2j}
   \right)}
   {\prod_{n=1}^\infty (1-zq^n)(1-z^{-1}q^{n-1})(1-q^n)}\ \ .
$$
We shall  use the symbol $\{ {b,j}\}^\wedge$ for these
irreducible representations of the affine algebra. The
formulas are easy to understand: they follow directly from
the representation (\ref{BFdec}) of the $\sl21$ current
algebra. In fact, each pair of fermionic fields contributes
a factor $\theta_1/\eta$ while the bosonic sl(2) and u(1)
current algebras are responsible for the characters
$\chi^{k-1}$ and an additional factor $\eta^{-1}$,
respectively.

\subsection{Atypical (class II) representations}

Nothing prevents us from evaluating the previous character
formulas at the points $b = j^\pm_m$. But the resulting functions
turn out to be the characters of indecomposable representations
$\{j^\pm_m,j\}^\wedge$ which contain one fermionic singular
multiplet. In order to state this more precisely, let us
consider in more detail the set of atypical labels,
$$ A \ :=\ \{\ \{j^\pm_m,j\}\ |\
      1/2 \,\leq\, j \,\leq\, k/2\  ;\  m \, \in\,
        \mathbb{Z}\ \}\ \ . 
$$
The set $A$ is visualized in Figure 2. Our picture shows clearly
that the projection to the $b$-coordinate of each element in $A$
is injective and hence it can be used to enumerate our atypical
labels. Note, however, that values $b \in (k+1)/2 \, \mathbb{Z}$
are omitted. This motivates to introduce an improved enumeration
map $\hat\nu$ from $A$ to non-zero half-integers which is defined
by
\begin{eqnarray*}
\hat\nu (\{j^+_m,j\}) \ = \ j^+_m - m  \ \ \mbox{ for } \ \ m\
\geq 0 \ \ & , & \ \ \hat\nu (\{j^-_m,j\}) \ = \ j^-_m - m + 1/2 \
\ \mbox{
for } \ \ m > 0 \\[2mm]
\hat\nu (\{j^-_m,j\}) \ = \ j^-_m + m  \ \ \mbox{ for } \ \ m\
\leq 0 \ \ & , & \ \ \hat\nu (\{j^+_m,j\}) \ = \ j^+_m + m - 1/2 \
\ \mbox{ for } \ \ m < 0\ \ .
\end{eqnarray*}
By construction, $\hat\nu$ is not only an injection but its image
now also consists of all nonzero half-integers. We may view $\hat
\nu$ as an affine version of the enumeration map $\nu(\{\pm j,j\})
= \pm j$ for representations of $\sl21$. \vspace{1cm}
\fig{\label{fig2}The set $A$ of atypical labels for the affine
$\sl21$ algebra. Even though the sl(2) spin $j$ is cut off at $j =
k/2$, there exist infinitely many atypical labels (black dots)
which are in one to one correspondence with the atypical labels of
the finite dimensional algebra $\sl21$ (central black and pink
dots). This correspondence is formalized by our map $\hat \nu$.}
{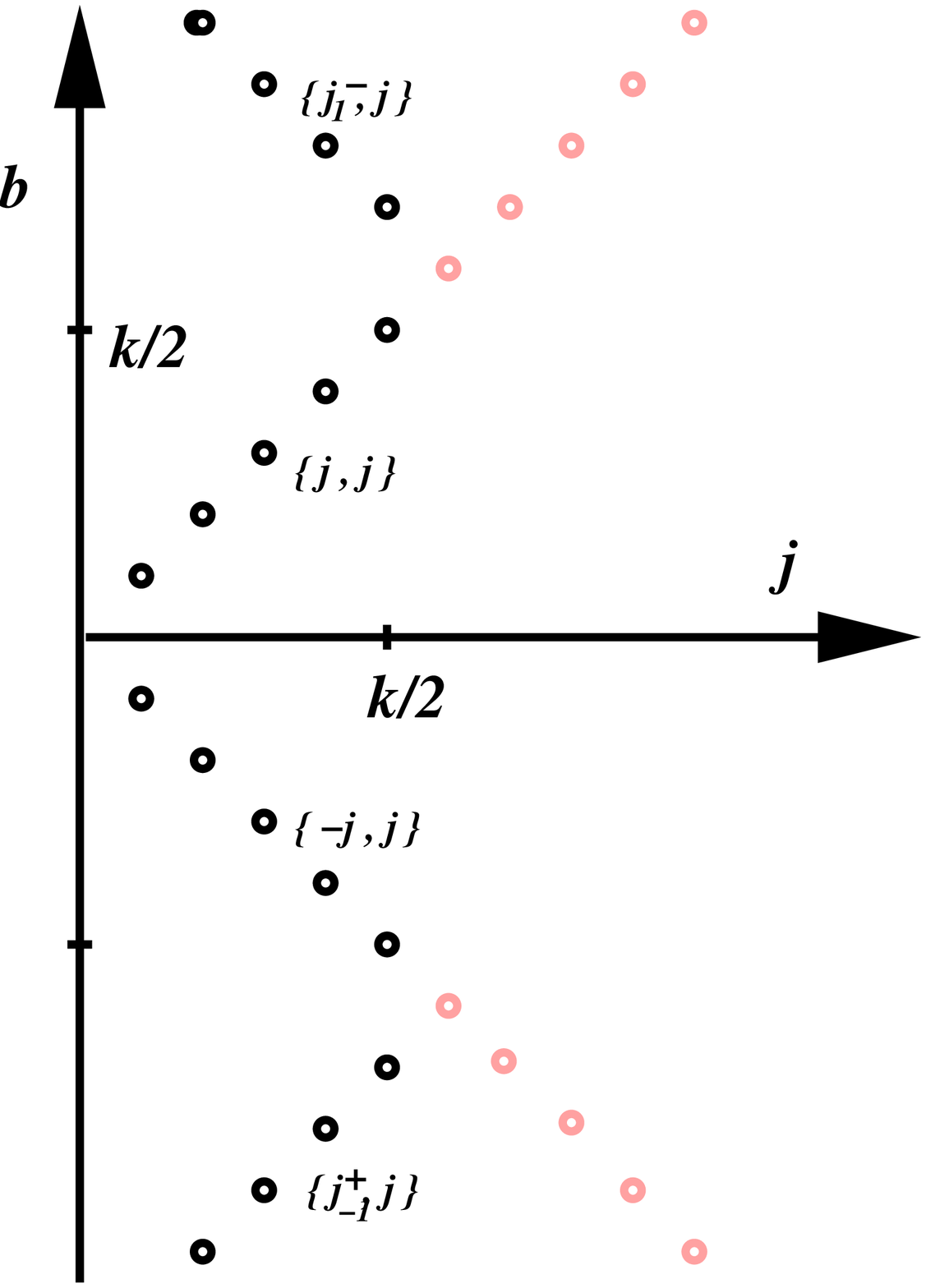}{9truecm} \figlabel{\basic} \vspace{5mm}
\smallskip

At first sight, the enumeration of atypical labels for our $\sl21$
current algebra may seem like a rather technical device. But there
is more to it. We recall that atypical labels $\{\pm j,j\}$ can
also be enumerated by non-zero half-integers, i.e.\ $\nu(\{\pm
j,j\}) = \pm j$. Our claim now is that the atypical class I
representation with label $\{ j^\pm_m,j\}$ behaves very similarly
to its finite dimensional counterpart $\nu^{-1} \circ \hat\nu (\{
j^\pm_m,j\})$ in the sense that
\begin{equation} \label{classI/II}
\chi^{I}_{\{ j^\pm_m,j\}}(q,z,\xi) \ = \
   \chi^{II}_{\{\hat\nu(\{j^\pm_m,j\})\}}(q,z,\xi)
  - \chi^{II}_{\{\hat\nu(\{\pm j_m,j\}) -
    1/2 {\text sgn}j^\pm_m\}}
  (q,z,\xi)\ \ .
\end{equation}
This formula is a rather central result for the representation
theory of our current algebra. Let us stress that it is the
affine version of a corresponding equality (\ref{chardiff})
between characters of $\sl21$ representations. As in the finite
dimensional setup, eq.\ (\ref{classI/II}) emerges from the
existence of fermionic singular vectors in atypical class I
representations. In the case $m=0$, it claims that the only
such singular vectors are those that appear in the atypical
Kac-module spanned by the ground states. When $m \neq 0$,
however, the ground states form a typical representation
and the singular vectors appear only on the $|m|^{th}$ level
of the class I module.
\smallskip

At first it may seem a bit surprising that affine
representations $\{j_m^\pm,j\}$ and $m\neq 0$ behave so
similarly to the Kac modules of $\sl21$. In the next
subsection we shall understand this behavior in terms
of spectral flow symmetries in the representation theory
of the current algebra. Before any study of spectral
flow automorphisms, it might be useful to illustrate
the similarity between atypical representations of the
current algebra and their
finite dimensional counterpart more explicitly, at least
for one example.  To this end, let us focus on the
representation $\{k/2+1,k/2\}^\wedge$ which we claim to
be a close cousin of the $\sl21$ representation $\{k/2+1/2,
k/2+1/2\}$. By construction, the ground states of the current
algebra representation transform in the typical multiplet
$\{k/2+1,k/2\}$ and they possess conformal weight $h = -1$.
From these vectors we can generate states with vanishing
conformal weight with the help of modes in the
current algebra. Such modes transform in the 8-dimensional
adjoint representation $\{0,1\}$ of $\sl21$. Through
decomposition of the tensor product between $\{k/2+1,k/2\}$
and $\{0,1\}$ one finds that the Verma module over
$\{k/2+1/2,k/2+1/2\}$ contains an atypical $\sl21$ multiplet
with conformal weight $h=0$. In fact, the results of
\cite{Gotz:2005jz} imply that the latter transforms according to
the projective cover $\mc{P}(k/2+1/2)$, see eq.\ (\ref{P}).
Not all of these states survive when we descend from the
Verma module to the class I representation. This step
involves removing bosonic singular vectors and a moment
of reflections shows that such vectors with $h=0$ exist
and that they transform in the submodule $\{k/2+1,k/2+1\}$
of $\mc{P}(k/2+1/2)$. Hence, the states with $h=0$ in our
class I representation decompose into the Kac-module
$\{k/2+1/2,k/2+1/2\}$ plus a bunch of typical representations.
The fermionic singular vectors that are responsible for 
eq.\ (\ref{classI/II}) transform in the subrepresentation
$\{k/2,k/2\}$ of $\{k/2+1/2,k/2+1/2\}$, giving rise to the
identity (\ref{classI/II}) with $j^-_1 = k/2+1$ and $j=k/2$.
\smallskip

Before we draw some conclusions from eq.\ (\ref{classI/II}), let
us quickly comment on our notations. Note that for $j=1/2$ and
$m = 0$ the above formula involves the character $\chi^{II}_{\{0\}}$
of a representation which has a somewhat special status. In fact, it
cannot be obtained as quotient of one of the indecomposable
representations $\{j^\pm_m,j\}^\wedge$, unlike the representations
with characters $\chi^{II}_{\{n/2\}}, n \neq 0,$. Instead, it arises
as a submodule of the representations $\{\pm 1/2,1/2\}^\wedge$.
Our discussion suggest that $\chi^{II}_{\{0\}}$ must be the
character of the vacuum representation. In the terminology of
Bowcock et al.\, the latter is a class IV representation. Thus,
we shall also write $\chi^{IV}_{\{0\}}$ for this quantity.
\smallskip

Even though equation (\ref{classI/II}) is not a closed formula for
the characters of class II representations, we can now use the
same trick as in section 2.1.2 and write characters of class II
representations as an infinite sum of class I characters,
\begin{equation}
\label{classIIchar}
 \chi^{II}_{\{\pm j\mp 1/2\}}(q,\xi,z) \ = \ - \sum_{n=0}^\infty
   \chi^I_{\hat\nu^{-1}(\pm j \pm n/2)}(q,\xi,z) \ \
\end{equation}
for $j =1/2,1,3/2, \dots$. Note that the map $\hat\nu$ is invertible
on all non-zero half-integers and it furnishes the label of the
Kac module that sits at the bottom of the corresponding class I
representation. By inserting our explicit formulas for class I
characters we find
\begin{eqnarray}
\chi^{II}_{\{j\}_\pm} (q,z,\xi) & = & - i\,  \frac{
   \theta_1(z^{1/2}\xi^{-1/2},q) \,
   \theta_1(z^{1/2}\xi^{1/2},q)} { \eta^{3}(q) \,
    \theta_1(z,q)}\ \times \\[2mm] & & \hspace*{1cm}
  \times \  \sum_{a \in \mathbb{Z}} \ q^{(k+1)a^2 \mp 2aj} \, \xi^{\pm j}\,
  \left(\frac{z^{-a(k+1) \pm j}}{1+q^a z^{-1/2}\xi^{-1/2}}
 -  \frac{z^{a(k+1) \mp j }} {1 + q^a z^{1/2}\xi^{-1/2}}
\right) \ \ . \nonumber
\end{eqnarray}
Character formulas of this type have to be used with some care:
Before the denominators are expanded, one should spit the
summation over $a$ into two parts. The one arising from positive
values of $a$ can be converted into a power series right away. In
all terms with non-negative $a$, however, one must first reduce
the fraction by $q^a$ such that the subsequent expansion contains
only non-negative powers of $q$. In the end, we recover the known
results on the representation theory of the affine $\sl21$ algebra
\cite{Bowcock:1997ce, Semikhatov:2003uc}. Our derivation was based
on three ingredients: the decoupling formulas (\ref{BFdec}) for
bosonic and fermionic generators, the structure (\ref{Kmdec}) of
atypical Kac modules for $\sl21$ and the fact that atypical class
I representations with $m\neq 0$ decompose in the same way as in
the case of $m=0$. We shall argue now that the last ingredient
emerges from spectral flow symmetries in the representation theory
of affine $\sl21$.

\subsection{Spectral flow symmetries}

The affine $\sl21$ algebra admits several interesting
automorphisms. We shall be mainly concerned with two such spectral
flow automorphisms $\gamma_\pm$. By construction, $\gamma_\pm$ are
defined on the entire current algebra, but for our purposes it is
sufficient to know how they act on the generators $B_0,H_0,L_0$,
$$ \gamma_\pm(B_0) \ = \ B_0 \pm k/2 \ \ \ , \ \ \ \
   \gamma_\pm(H_0) \ = \ H_0 + k/2 \ \ \ , \ \ \
   \gamma_\pm(L_0) \ = \ L_0 + H_0 \mp B_0 \ .
$$
From these formulas we may infer how (super-)characters behave
under the action of $\gamma_\pm$ and this in turn is sufficient to
determine how spectral flow automorphisms map representations of
the current algebra onto each other. Along with $\gamma_\pm$ we
shall also be interested in the composite automorphism $\gamma =
\gamma_+ \gamma_-^{-1}$ which acts as
$$ \gamma(B_0) \ = \ B_0 + k \ \ \ , \ \ \ \
   \gamma(H_0) \ = \ H_0 \ \ \ , \ \ \
   \gamma(L_0) \ = \ L_0 - 2 B_0 - k \ \ .
$$
Any automorphisms of the current algebra gives rise to a map
between representations and hence to a map between characters.
From the action on the zero modes $B_0, H_0$ and $L_0$ we can
easily read off that
\begin{equation}\label{gchi}
\gamma_\pm \chi(q,\xi,z) \ = \ \xi^{\pm k/2} z^{k/2}
                 \,  \chi(q,q^{\mp 1}\xi, qz)
\ \ \ , \ \ \ \gamma \chi(q,\xi,z) \ = \
  \xi^k q^{-k}\,  \chi(q,q^{-2}\xi,z) \ \
\end{equation}
for all characters $\chi$ of the $\sl21$ current algebra. If
${\cal R}$ is any representation of $\asl21$ and $\chi_{\cal R}$
is its character, then the image $\sigma {\cal R}$ of ${\cal R}$
under an automorphism $\sigma$ obeys
$$ \sigma \chi_{\cal R} (q,\xi,z) \ = \ \chi_{\sigma
{\cal R}} (q,\xi,z)\ \ . $$  Given the character $\chi_{\cal R}$
of some representation ${\cal R}$, we can use eqs.\ (\ref{gchi})
to compute its image under the above automorphisms $\sigma =
\gamma_\pm, \gamma$. This in turn allows us to recover uniquely
the representations $\gamma_\pm {\cal R}$ and $\gamma {\cal R}$.
\smallskip

In the following we shall spell out the action of our spectral
flow automorphisms on the class I and II representations we have
studied above. Our rather compact notations allow us to summarize
the results for the spectral flow automorphisms $\gamma_\pm$ in a
single line
\begin{equation}
\gamma_\pm(\{ {b,j}\}^\wedge) \ = \ \{b\pm k/2 \pm \oh, k/2 +
    \oh - j\}^\wedge\ \ , \ \
\gamma_\pm(\{n/2\}^\wedge) \ = \ \{n/2\pm k/2\}^\wedge\ \ .
\end{equation}
To verify our assertions, the reader is invited to convert them
into identities between super-characters and to check these
identities by direct computation. The formulas become somewhat
more explicit if we label irreducible representations according to
the representation their ground states transform in,
\begin{equation} \begin{split}
\{ b,j\} & \stackrel{\gamma_\pm}{\longrightarrow} \
    \{b\pm k/2\pm \oh,k/2 + \oh -j\}
 \ \ \ \ \ \ \ \mbox{ for } \ \ b \neq \pm (j - k - 1)\\[2mm]
\{ b,j\} & \stackrel{\gamma_\pm}{\longrightarrow} \
    \{k/2 +\oh - j \}_\mp
 \hspace*{2.5cm}\ \ \ \mbox{ for } \ \ b = \pm(j - k - 1)\\[2mm]
\{j\}_\pm & \stackrel{\gamma_\pm}{\longrightarrow} \
    \{\pm(j + k/2 + \oh),k/2 + \oh -j\} \ \ \ \mbox{ for } \ \
  j \neq 0 \\[2mm]
\{j\}_\mp & \stackrel{\gamma_\pm}{\longrightarrow} \
    \{k/2  -j\}_\pm \ \ \ \ \ \ , \ \ \ \ \ \
\{0\} \ \stackrel{\gamma_\pm}{\longrightarrow} \
\{ k/2\}_\pm \ \ .\end{split}
\end{equation}
The third line, for example, tells us that the image of the
irreducible representation generated from the atypical
representation $\{j\}_+$ under the action of $\gamma_+$ is
an irreducible representation whose ground states transform
in the typical representation $\{ j + k/2 + \oh,k/2 + \oh -j\}$.
The latter may be obtained from the corresponding Verma module
by removing singular vectors on some excited levels.
\smallskip

We also want to spell out analogous formulas for the automorphism
$\gamma = \gamma_+ \circ \gamma_-^{-1}$. In the compact notation,
its action is given by
\begin{equation}
\gamma(\{ {b,j}\}^\wedge) \ = \ \{b + k + 1, j\}^\wedge\ \ , \ \
\gamma(\{n/2\}^\wedge) \ = \ \{n/2 + k\}^\wedge\ \ .
\end{equation}
Note that the symmetry $\gamma$ maps sectors whose ground states
transform in an atypical representation $\{j\}_\pm$ of the Lie
superalgebra $\sl21$ into sectors with typical spaces of ground
states according to the following rules,
\begin{eqnarray*}
\{j\}_\pm & \stackrel{\gamma^m}{\longrightarrow} &
   \{j_m^\pm,j\}    \ \ \ \mbox{ for }
 \ \ \ \pm m \ > \ 0 \\[2mm]
 \{j\}_\pm & \stackrel{\gamma^m}{\longrightarrow} &
  \{j_m^\pm \pm 1/2,j + 1/2\}
  \ \ \ \mbox{ for }  \ \ \ \pm m \ < \ 0 \ \ .
\end{eqnarray*}
Hence, the existence of the spectral flow symmetries explains
why the representations $\{j_m^\pm,j\}$ behave like atypical
representations of the affine $\sl21$ algebra: they are simply
related to the sectors erected over atypical $\sl21$
representations by an automorphism.

\subsection{Modular transformation and S-matrix}

We would like to conclude this section on the representation
theory of the $\sl21$ current algebra with a few comments on
modular properties of the characters. In the following we shall
consider the characters as functions of $\tau, \sigma$ and $\nu$.
They are related to the variables we used above through $q= \exp
2\pi i \tau, z=\exp 2\pi i\sigma$ and $\xi=\exp 2\pi i \nu$, as
usual. From our explicit formula (\ref{typeI}) for characters of
class I representations it is easy to infer the auxiliary formula
$$
     \chi_{\{b=0,j\}}^{I}(-\frac{1}{\tau},\frac{\sigma}{\tau},\frac{\nu}{\tau})\ =\ -
     \frac{1}{\sqrt{-i\tau}}
     e^{\frac{i\pi k}{ 2\tau}
     \sigma^{2}}e^{\frac{i\pi}{
     2\tau}\nu^{2}}\sum_{j' =1/2}^{\frac{k}{2}}\sqrt\frac{2}{k+1}\sin
    \frac{4\pi jj'}{k+1}\chi^{I}_{\{b=0,j'\}}(\tau,\sigma,\nu)\ . 
$$
Note that the right hand side contains an explicit $\tau$
dependence which, if we demand that the modular transform be
interpreted in a conventional sense and closes onto characters,
suggests the contribution of a continuous spectrum of exponents.
The need is confirmed by the modular transformation of the
character for $\{b,j\}$ representations with $b\neq 0$, which
require an integral representation of $e^{2i\pi\tau b^{2}/(k+1)}$
etc. After some Gaussian integration, we find
\begin{eqnarray}
    \chi_{\{b,j\}}^{I}(-\frac{1}{\tau},\frac{\sigma}{\tau},
     \frac{\nu}{\tau})\ =\ i
        e^{\frac{i\pi k}{2\tau}
        (\sigma^{2}-\nu^{2})}
        \sum_{j' =1/2}^{\frac{k}{2}}\frac{2}{k+1}\sin
       \frac{4\pi jj'}{k+1}\int_{-\infty}^{\infty}db'\, 
     e^\frac{4i\pi bb'}{k+1}
       \chi^{I}_{\{b',j'\}}(\tau,\sigma,\nu)
       \end{eqnarray}
where we formally evaluated integrals of the type $\int \exp (
-i\tau x^{2})=\sqrt{\pi/i\tau}$ [of course, the integrals are
naively divergent as $\hbox{Im }\tau>0$.]
\smallskip

Modular transformations of the type II and IV characters are a bit
more cumbersome to work out. It can be attacked rather efficiently
using our representations (\ref{classIIchar}) as infinite sums of
class I characters. Here we shall content ourselves with the
example of the class IV representation at $k=1$. If we also set
$\sigma=\nu=0$ we find that
\begin{equation}
    \chi^{IV}_{\{0\}}(-1/\tau)\ =\ -\frac{i}{2}\int_{-\infty}^{\infty}
    \ \frac{db'}{\cos\pi b'}\ \chi^{I}_{\{b',1/2\}}(\tau)
    \end{equation}
where the contour has to avoid the poles. Rotating into the purely
imaginary direction gives a contribution from poles which is
easily identified with $\chi^{IV}_{\{0\}}$. The remaining integral 
can be factored in terms of $\chi^{I}_{\{0,1/2\}}$,
\begin{equation}
    \chi^{IV}_{\{0\}}(-1/\tau)\ =\ -\chi^{IV}_{\{0\}}(\tau)\, +\, \int\  d\alpha
   \ \frac{q^{\frac{\alpha^{2}}{{2}}}}{\cosh\pi\alpha}
  \  \chi^{I}_{\{0,1/2\}}(\tau)\ \ .
   \end{equation}
 We thus recover by this very elementary means the results of 
 \cite{Bowcock:1997ce}
 obtained through use of the Mordell integral \cite{Mordell}. The construction of modular
 invariants using these transformation formulas is a complex problem, which
 we shall address later in the case $k=1$.

\section{The state space and partition functions}

Our aim now is to formulate a proposal for the states space of the
$\sl21$ WZNW model. We shall then verify our claim in the special
case $k=1$ through a free field representation of the model. The
third subsection is devoted to the partition function of the
theory. The latter forgets all information about the complicated
way in which irreducible blocks are glued together to build ${\cal
J}$. We then specialize once more to $k=1$ and comment on the 
global topology of the target space. 

\subsection{The proposal for integer level $k$}

It is now rather straightforward to come up with a proposal for
the state space of the WZNW model on $\SL21$. In fact, we can
simply depart from formula (\ref{LRdec}) and make it symmetric
with respect to the action of our spectral flow symmetry. The
invariance under the action of $\gamma$ should be considered as an
additional input. In principle, the spectral flow symmetry of the
$\sl21$ current algebra could be broken by the physical couplings
of the theory. Since this did not happen for the $\GL11$ WZNW
model, it seems natural to propose
\begin{equation} \label{HCFT}
 {\cal H}_{CFT}\ = \ \sum_{j=1/2}^{k/2} \sum_{b\neq j^\pm_m} \
    \{b,j\}^\wedge_L \otimes \{-b,j\}^\wedge_R \ \oplus \
  {\cal J}^\wedge
\end{equation}
where ${\cal J}^\wedge$ is a single indecomposable representation of the
two (anti-)commuting super current algebras that contains all the
atypical contributions. It is composed from the atypical building
blocks $\{l_1\} \otimes \{l_2\}$ in the same way as in the
minisuperspace theory. To obtain the corresponding diagram one
simply has to replace $\{j\}_\pm = \{\pm j\}$ with $\{\pm
j\}^\wedge$.
\smallskip

By construction, all the $\sl21$ currents act on the state space
(\ref{HCFT}) and they obey periodic boundary conditions. This
applies in particular to the fermionic fields. One can find a
second, closely related theory in which only bosonic fields are
periodic. In order to construct its state space, we need to
revisit our discussion of spectral flow symmetries. As we have
mentioned above, the automorphisms we have investigated in the
previous section all extend to the entire current algebra. In
particular, they map fermionic modes with integer mode numbers
onto each other, i.e.\ they respect periodic boundary conditions
on the fermionic currents. There exists yet another important
isomorphism that intertwines between integer and half-integer mode
numbers for the fermionic generators. It can be considered as the
square root of the automorphism $\gamma$. On the bosonic zero
modes, the new isomorphism $\vartheta$ is given by
$$ \vartheta(B_0) \ = \ B_0 + k/2 \ \ \ , \ \ \ \
   \vartheta(H_0) \ = \ H_0 \ \ \ , \ \ \
   \vartheta(L_0) \ = \ L_0 -  B_0 - k/2 \ \ .
$$
$\vartheta$ extends to the full current algebra such that it acts
trivially on the bosonic sl(2) currents and it shifts modes of the
fermionic currents by $\pm 1/2$, as usual.
\smallskip

Our isomorphism $\vartheta$ induces a map between representations
of the current algebra with integer fermionic modes and a new type
of representations in which fermionic generators have half integer
mode numbers. According to the usual terminology, the former class
of representations form the R sector while the latter belong to
the NS sector of the theory. The theory with state space
(\ref{HCFT}) includes exclusively R sector representations in
which all currents obey periodic boundary conditions. Another
option is to consider a theory that encompasses both R and
NS sector with the state space given by
$$ \tilde {\cal H}_{CFT} \ = \ {\cal H}_{CFT}^{R} \oplus
  {\cal H}_{CFT}^{NS} \ = \ {\cal H}_{CFT} \oplus
   \vartheta {\cal H}_{CFT} \ \ . $$
Note that the NS sector has exactly the same intricate structure
as the R sector since the former is the image of the latter under 
the action of $\theta$. In the following we shall refer to both 
models as WZNW model on $\SL21$. Even though it seems natural to 
include the NS sector, it is not required by all applications.

\subsection{Free field representation at $k=1$}

So far, the main motivation for our proposal (\ref{HCFT}) came
from the harmonic analysis on $\SL21$. By construction, we are
guaranteed to recover the correct state space of the particle
limit when we send the level $k$ to infinity. Our formula
(\ref{HCFT}) applies to finite $k$ and it suggest that field
theory effects would merely truncate the spin $j$ to an value $j
\leq k/2$ and then make the whole theory symmetric with respect to
spectral flow. We are now going to test this proposal in the
extreme quantum case, namely at $k=1$. At this point, the WZNW
model admits a free field representation that we are going to
spell out momentarily. It involves a pair of symplectic fermions
$\eta_{1},\eta_{2}$, and a pair of bosons $\phi,\phi'$. While the
boson $\phi$ comes with the usual metric, $\phi'$ is assumed to be
time-like. For their propagators this means
\begin{eqnarray}
    \langle\, \eta_{1}(z,\bar{z})\, \eta_{2}(w,\bar{w})\, \rangle &=&
     -\ln|z-w|^{2}\nonumber\\
    \langle\ \phi(z,\bar{z})\ \phi(w,\bar{w})\ \rangle&=&
     -\ln|z-w|^2\nonumber\\
    \langle\, \phi'(z,\bar{z})\, \phi'(w,\bar{w})\, \rangle&=&
    \ln|z-w|^2\ \ . 
\end{eqnarray}
Note that the central charge of this free field theory is
$c=-2+1+1=0$ and hence it agrees with the central charge of
$\SL21$ WZNW models. We shall begin our discussion of the WZNW
model with explicit formulas for the currents. In order to
construct the four bosonic currents, we need to split the 
space-like bosonic field $\phi(z,\bar{z}) = \varphi(z) +
\bar{\varphi}(\bar{z})$ into its chiral components. Our
bosonic currents then read,
\begin{eqnarray}
    E^{+}(z)\ = \ e^{\sqrt{2}i\varphi(z)} \ \ \ & , & \ \ \
    E^{-}(z) \ = \ e^{-\sqrt{2}i\varphi(z)}\nonumber\\[2mm]
    H (z)\ = \ \frac{1}{\sqrt{2}}\, i\partial\phi(z) \ \ \ & , & \ \ \
    B (z)\ = \ -\frac{1}{\sqrt{2}}\, i\partial\phi'(z)\ \ .
    \end{eqnarray}
The necessity to split $\phi$ into its chiral components means 
that the boson $\phi$ is compactified to the so-called self-dual 
radius, as usual in
the free field representation of the SU(2) WZNW model at
level $k=1$. In addition, the following expressions for the
four fermionic currents also involve the chiral components
$\varphi'$ and $\bar \varphi'$ of the time-like bosonic field
$\phi'(z,\bar z) = \varphi'(z) + \bar \varphi'(\bar z)$,
\begin{equation} \label{Fcurrents}
    V^{\pm}(z) \ =\ e ^{\frac{1}{\sqrt{2}}i(\pm
    \varphi(z)+\varphi'(z))}\partial\eta_{1}(z)\ \ \ , \ \ \
    W^{\pm}(z) \ = \ e^{\frac{1}{\sqrt{2}}i(\pm
     \varphi(z)-\varphi'(z))}\partial\eta_{2}(z)\ \ . 
\end{equation}
Similarly, one may spell out the anti-holomorphic generators of
the $\sl21$ current algebra. It is rather easy to check that the
above expressions give rise to fields with the correct operator
product expansions. Let us note that the free field representation
we consider in this section has to be distinguished clearly from
the Kac-Wakimoto type construction (\ref{BFdec}) we have used
earlier to construct the characters at integer levels $k$. We
shall comment on this a bit more later on.
\smallskip

It is possible to check that fields of dimension zero can be
organized exactly as it is suggested by our diagram
(\ref{diagram}). We shall just sketch the relevant arguments because
a full proof is rather laborious to write down. Let us consider
the left part of the diagram only and identify the fields that
make up the various blocks of the composition series. Clearly, the
$\{0\}\times \{0\}$ representation at the top corresponds to the
field $\eta_{1}\eta_{2}$. From here we can act with the fermionic
currents $W^\pm, \bar V^\pm$ and arrive at expressions for the two
blocks on the intermediate level of the diagram,
\begin{eqnarray}
   \{1/2\}_{-}\times \{0\}\ :  \ \ & &  e^{\pm
    i\varphi/\sqrt{2}}\, e^{-i\varphi'/\sqrt{2}}\, \eta_2 \ \ \ , \ \ \
    e^{-i\sqrt{2}\varphi'}\ \eta_2 \partial\eta_{2} \ \ ,
    \nonumber\\[2mm]
    \{0\}\times\{1/2\}_{+}\ : \ \ & &  e^{\pm
    i\bar{\varphi}/\sqrt{2}}\, e^{i\bar{\varphi}'/\sqrt{2}}\,
    \eta_{1}\ \ \ , \ \ \
    e^{i\sqrt{2}\bar{\varphi}'}\, \eta_{1}\bar{\partial}\eta_{1}
   \ \ .  \nonumber
    \end{eqnarray}
From the previous formulas we can read off the fields that make up
the topmost representation $\{1/2\}_{-}\times \{1/2\}_{+}$ in our
diagram,
\begin{equation}
 \{1/2\}_{-}\times
      \{1/2\}_{+} \ : \ \ \ \ \ \ \ \left\{\begin{array}{c}
    e^{\pm i\varphi/\sqrt{2}}\, e^{-i\varphi'/\sqrt{2}}\,
    \eta_{2}\\[1mm]
    e^{-i\sqrt{2}\varphi'}\, \eta_{2}\partial\eta_{2}\end{array}
   \right. \ \times \
   \left\{\begin{array}{c}
      e^{\pm i\bar{\varphi}/\sqrt{2}}\, e^{i\bar{\varphi}'/\sqrt{2}}\,
       \eta_{1}\\[1mm]
      e^{i\sqrt{2}\bar{\varphi}'}\, \eta_{1}\bar{\partial}\eta_{1}
      \ \ . \end{array}\right.  \label{bigie}
\end{equation}
Acting with the holomorphic fermionic currents $V^\pm(z)$ we
arrive at the following formulas for fields that belong to the
multiplet
\begin{equation}
  \{1\}_{-}\times \{1/2\}_{+}\ : \ \ \ \ \   \left\{\begin{array}{c}
    (\pm \partial\phi+\partial\phi')\, \eta_{2}\partial\eta_{2}
    \, e^{\pm i\varphi/\sqrt{2}}\, e^{-3i\varphi'/\sqrt{2}}\\[1mm]
\eta_{2}\, e^{\pm i\sqrt{2}\varphi}\,
e^{-i\sqrt{2}\varphi'}\\[1mm]
\eta_{2}\partial\phi \, e^{-i\sqrt{2}\varphi'}
\end{array}
       \right.\ \times\
       \left\{\begin{array}{c}
      e^{\pm i\bar{\varphi}/\sqrt{2}}\,
      e^{i\bar{\varphi}'/\sqrt{2}}\, \\[1mm]
      e^{i\sqrt{2}\bar{\varphi}'}\, \bar{\partial}\eta_{1}
      \ \ .. \end{array}\right.
\end{equation}
on the intermediate level of the diagram. Our notation means that
every product of the three holomorphic fields on the left hand
side with the three anti-holomorphic fields on the right hand side
is part of this 9-dimensional block. Similarly, we can now descend
to the bottom of the diagram,
\begin{equation}
    \{1/2\}_{-}\times \{1/2\}_{+} \ : \ \ \ \ \ \ \  \left\{\begin{array}{c}
      e^{\pm i\varphi/\sqrt{2}}\, e^{-i\varphi'/\sqrt{2}}\\[1mm]
      \eta_{2}\partial \phi'\,
      e^{-i\sqrt{2}\varphi'} \end{array}\right.\ \times\
       \left\{\begin{array}{c}
      e^{\pm i\bar{\varphi}/\sqrt{2}}\,
      e^{i\bar{\varphi}'/\sqrt{2}}\, \\[1mm]
      e^{i\sqrt{2}\bar{\varphi}'}\,  \bar{\partial}\eta_{1}
      \end{array}\right.
\end{equation}
Finally, the representation $\{0\}\times \{0\}$ in center bottom
position is represented by the identity field. It is easy but
laborious to check that the different representations are
connected by the action of the left and right generators as
indicated in the diagram. In checking this, notice that
$\eta_{2}\partial\phi' e^{-i\sqrt{2}\varphi'}\equiv
\partial\eta_{2}e^{-i\sqrt{2}\varphi'}$ up to a total
derivative.
\smallskip

There are a number of interesting further comments and observations
that we would like to make. Let us begin with a brief comment on
the relation with Kac-Wakimoto like representations of the form
(\ref{BFdec}). As discussed in \cite{Gotz:2006qp}, a naive evaluation
of the action of the latter on vertex operators leads to a
much simpler picture in which the atypical sector is a smooth
deformation of the typical part. In particular, there is no
mixing between left- and right movers as in the case of the
representation ${\cal J}^\wedge$. In order to see the latter,
the screening charge of the Kac-Wakimoto representation must
be taken into account (see \cite{Gotz:2006qp} for details). The free
field representation we have employed in this subsection is much
simpler to use, but it is restricted to $k=1$.
\smallskip

The free field representation also allows us
to illustrate very explicitly how atypical fields of dimension $h=0$
are embedded into sectors with ground states in typical multiplets once
their spin exceeds $k/2$.
Take, for instance, the field
$O=e^{i\sqrt{2}(\varphi-\varphi')}\eta_{2}$ from the $\{1\}_{-}$
representation and observe that
\begin{equation}
    E^{-}(z)\  e^{i\sqrt{2}(\varphi-\varphi')}\, \eta_{2}(w)\ =\
    \frac{1}{ (z-w)^{2}} \ :e^{-i\sqrt{2}\phi(z)}\,
    e^{i\sqrt{2}(\varphi-\varphi')}\, \eta_{2}(w): \ + \dots \ \ .
    \end{equation}
Thus $O$ is not a highest weight of the current algebra, and
applying a current operator can give rise to a field of dimension
$h=-1$, namely the field $P = e^{-i\sqrt{2}\varphi'}\eta_{2}$.
This field belongs to the typical representation $\{-3/2,1/2\}$ in
the sector $\{1\}_-$ with affine highest weight $Q =
e^{i(\varphi-3\varphi')/\sqrt{2}}\eta_{2}\partial\eta_{2}$. One may easily
generalize these observations to all representations in the
complex, hence confirming our analysis based on spectral flow.
\smallskip

Let us finally come to the most important point, which concerns
the possible construction of consistent theories\footnote{
Consistency in this paragraph refers to the existence of 
genus zero correlators obeying the usual factorization 
constraints. The construction and behavior of torus 
amplitudes is not addressed.} that are realized on a subspace 
of our state space (\ref{HCFT}). Note that ${\cal H}$ does 
contain a large number of fermionic singular vectors that
we decided not do decouple, partly because the minisuperspace
analysis suggested that it was unnatural to do so. But to a
certain extend one should consider (\ref{HCFT}) as some 
kind of maximal choice from which other models can be obtained by
consistent decoupling of singular vectors. In general there can be
several such reduced theories. Once more we may use our free field
representation for the $k=1$ model to illustrate nicely how this
works. Note that the expressions (\ref{Fcurrents}) for the
currents only contain derivatives of the fermionic fields. Hence,
we do not spoil the $\sl21$ current algebra symmetry if we decide
to work with a model in which the fundamental fields are e.g.\
$\phi, \phi', \eta_2$ and $\partial \eta_1$. Since $\eta_1$ is not
part of this model, some of the sectors we discussed above do no
longer appear. This concerns the sectors $\{0\} \times \{0\}$ and
$\{1/2\}_-\times \{1/2\}_+$ on the top floor of ${\cal J}^\wedge$
and the sector $\{0\} \times \{1/2\}_+$ on the intermediate floor.
In the resulting model, there is still a single atypical sector
that comprises all the irreducible atypicals, but it is reduced to
two floors and has the shape of a saw blade. Obviously, a similar
analysis applies to the theory that contains $\partial \eta_2$
instead of $\eta_2$. But we can even go one step further and drop
both $\eta_1$ and $\eta_2$ so that only fermionic derivatives
remain. What results is a model whose atypical sector decomposes
into an infinite sum of irreducibles. The latter are the sectors
that appear on the bottom floor of ${\cal J}^\wedge$, i.e.\ $\{0\}
\times \{0\}$ and $\{1/2\}_- \times \{1/2\}_+$ from our list
above. All others need either $\eta_1$ or $\eta_2$. Similar
phenomena are possible at other levels. We shall see another
explicit example in section 5.2. Let us stress, however, that 
the free field construction at $k=1$ does not include the 
``defining'' $\{0,1/2\}$ field of the WZNW model. A careful 
study of the Knizhnik-Zamolodchikov equations 
\cite{Maassarani:1996jn,Bhaseen:2000mi}
shows that 
consistency in the presence of the $\{0,1/2\}$ sector 
requires the identity field to be embedded into an 
indecomposable sector. In this sense, the fully truncated 
atypical sector we have just described cannot be embedded 
into the $\sl21$ WZNW model.

\subsection{Partition functions}

We now go back to the full theory based on our proposal (\ref{HCFT}). 
We would like to compute
the partition function of the model, with and without inclusion of
the NS sector. Since partition functions are obtained by taking
the trace over the state space, the details of the action of
fermionic generators in the atypical sector ${\cal J}^\wedge$ do
not show up in the result. In other words, the contribution from
the indecomposable ${\cal J}^\wedge$ is the same as if we would
take the trace over a sum of its irreducible components. The
latter can be resumed as follows,
\begin{eqnarray*}
\text{str}_{{\cal J}^\wedge}\  q^{L_0 + \bar L_0} & = &
 \sum_{\nu \in \mathbb{Z}/2} \, 2 \chi^{II}_{\{\nu\}}(\bar q)\,
      \chi^{II}_{\{-\nu\}}(q) - \chi^{II}_{\{\nu+1/2\}}(\bar q)\,
      \chi^{II}_{\{-\nu\}}(q) - \chi^{II}_{\{\nu-1/2\}}(\bar q) \,
      \chi^{II}_{\{-\nu\}}(q)   \\[2mm]
& = & \sum_{\nu \in \mathbb{Z}/2} \, \left( \chi^{II}_{\{\nu\}}(\bar q)
      - \chi^{II}_{\{\nu-1/2\}}(\bar q)\right) \left(
       \chi^{II}_{\{-\nu\}}(q) -
      \chi^{II}_{\{-\nu+1/2\}}(q)\right) \\[2mm]
& = & \sum_{\nu = 1/2,1,\dots} \, \left( \chi^{II}_{\{\nu\}}(\bar q)
      - \chi^{II}_{\{\nu-1/2\}}(\bar q)\right) \left(
       \chi^{II}_{\{-\nu\}}(q) -
      \chi^{II}_{\{-\nu+1/2\}}(q)\right) \\[2mm]
& & + \  \sum_{\nu = 1/2,1,\dots} \, \left( \chi^{II}_{\{-\nu\}}(\bar q)
      - \chi^{II}_{\{-\nu+1/2\}}(\bar q)\right) \left(
       \chi^{II}_{\{\nu\}}(q) -
      \chi^{II}_{\{\nu-1/2\}}(q)\right) \\[2mm]
& = & \sum_\pm \sum_{m \in \mathbb{Z}} \sum_{j=1/2}^{k/2} \,
   \chi^I_{\{j^\pm_m,j\}}(\bar q)\, \chi^I_{\{-j^\pm_m,j\}}(q)\ \ . 
\end{eqnarray*}
In the last step we have inserted the relation (\ref{classI/II})
between characters of class I and class II representations and we
used the isomorphism $\hat \nu$ to convert the sum over non-zero
half-integers into a sum over $m$ and $j$. Our result shows that
the contribution from the atypical representations agrees exactly
with the part that we omitted from the typical sector of the
theory. Hence, the full partition function becomes
\begin{equation}\label{PFH}
Z(q) \ = \ \sum_{j=1/2}^{k/2} \sum_{b\in \mathbb{R}}
    \chi^I_{\{b,j\}}(\bar q) \chi^I_{\{-b,j\}}(q)\ \ .
\end{equation}
Let us also briefly discuss how the partition functions is
modified when we want to include the NS sector. In that case, the
trace extends over both ${\cal H}_{CFT}$ and its image under the
spectral flow $\vartheta$. The modular invariant partition
function $\tilde Z(q)$ of this theory contains four different
terms, two in which $(-1)^F$ is inserted and two in which it is
not. It is customary to label the corresponding contributions with
$R$, $sR$, $NS$ and $sNS$ where the small $s$ signals the
insertion of $(-1)^F$. With these notations, the standard
super-characters we have discussed throughout the previous section
should all carry a superscript $sR$. It is easy to find explicit
formulas for the other three sets of (super-)characters using the
relation
$$
\chi^{sNS}(q,\xi,z) \ = \ \vartheta \chi^{sR}(q,\xi,z) \ = \
\xi^{k/2} q^{-k/2} \chi^{sR}(q,q^{-1} \xi, z) \ \
$$
to convert $sR$ characters into $sNS$ super-characters. The same
prescription is used when we construct $NS$ characters from the R
sector, only that we have to replace the $R$ super-characters
$\chi$ by ordinary characters. The partition function $\tilde
Z(q)$, finally, has the same form as eq.\ (\ref{PFH}) with an
additional summation over all four types of terms.
\medskip

Let us illustrate the previous results in the case of $k=1$ again.
The ($sR$) characters of this theory take a particularly simple
form, as was first observed by Bowcock et al.\ in
\cite{Bowcock:1997ce}, 
\begin{equation}
    \chi_{\{b,1/2\}}^{I}(q,z,\xi)\ =\ \frac{q^{-b^{2}/2}\xi^{b}}{\eta(q)}
    \left[\chi_{1/2}(q,\xi)\chi_{0}(q,z)-\chi_{0}(q,\xi)\chi_{1/2}(q,z)\right]
    \end{equation}
where $\chi_{0}$ and $\chi_{1/2}$ are  $SU(2)$ level one
characters for spin $0$ and $1/2$ respectively. This expression
allows us to determine the modular invariant physical partition
function $\tilde Z$ involving periodic or antiperiodic boundary
conditions for the fermions along both periods of the torus. The
doubly periodic sector ($\chi^{sR}$) gives a vanishing
contribution for characters $\chi_{\{b,1/2\}}$ since the
super-dimension of the horizontal Kac modules vanishes. We are
left with three contributions, which read respectively
\begin{eqnarray}
    \chi^{R}_{\{b,1/2\}}(q)&=&2\frac{q^{-b^{2}/2}}{\eta(q)}\
    \chi_{1/2}(q)\chi_{0}(q)\\
    \chi^{NS}_{\{b,1/2\}}(q)&=&\frac{q^{-b^{2}/2}}{
    \eta(q)}\ \left[\chi_{0}^{2}(q)+\chi_{1/2}^{2}(q)\right]\\
    \chi^{sNS}_{\{b,1/2\}}(q)&=&\frac{q^{-b^{2}/2}}{
    \eta(q)}\ \left[\chi_{0}^{2}(q)-\chi_{1/2}^{2}(q)\right]\ \ .
    \end{eqnarray}
Their modular transformations are easy to obtain for $b=0$
\begin{eqnarray}
    \chi^{R}_{\{0,1/2\}}(-1/\tau)&=&\frac{1}{
    \sqrt{-i\tau}}\ \chi^{sNS}_{\{0,1/2\}}(\tau)\nonumber\\
    \chi^{NS}_{\{0,1/2\}}(-1/\tau)&=&\frac{1}{
    \sqrt{-i\tau}}\ \chi^{NS}_{\{0,1/2\}}(\tau)\\
    \chi^{sNS}_{\{0,1/2\}}(-1/\tau)&=&\frac{1}{
        \sqrt{-i\tau}}\ \chi^{R}_{\{0,1/2\}}(\tau)\ \ .
     \nonumber   \end{eqnarray}
Obviously, the $\tau$ dependence of these formulas originates in
the $1/\eta$ term in the characters, and has to be compensated by
a similar factor coming from the $b$ sum in order to obtain a
modular invariant quantity.
\smallskip

The question we want to ask now is what kind of sum over $b$ one
should consider. It seems at first sight that, since we are
dealing with $\SL21$, the $b$ number should be discrete, in
agreement with the imaginary exponential appearing in section 2.2
(note that this is also compatible with invariance under action of
the spectral flow). It is likely that such a theory would make
sense in genus zero. Difficulties arise, however, when we try to
establish modular invariance of the partition function, i.e.\
consider the theory in genus one. Indeed, even if $b$ is
discretized, there is no reason to truncate its range, and thus
the naive spectrum of conformal weights is unbounded from below,
and exhibits arbitrarily large negative dimensions. This should
not come as a surprise since the metric on the group is not
positive definite, and thus the naive functional integral in the
WZNW model is divergent. What is required to obtain a physical
partition function on the torus - one that could be compared with
the spectrum of some lattice Hamiltonian say - is some sort of
analytic continuation. \smallskip

This raises some interesting questions on which we would like to
digress briefly. For a compactified  time-like boson, partition
functions would involve sums of the form
$$
\sum_{n}\ e^{\lambda n^{2}},~~~\hbox{Re }\lambda>0\ \ . 
$$
This sum is obviously divergent, but one could be tempted to give
it a meaning by analytical continuation from a similar sum with
$\hbox{Re }\lambda<0$. An equivalent problem arises when we try to
continue a theta function such as
$$
    \theta(\tau)\ =\ \sum_{n}e^{i\pi\tau n^{2}}
$$
into the lower half plane $\hbox{Im }\tau<0$. It is known that
this continuation is not possible, since the $\theta$ function has
singularities which are {\it dense} on the real axis (a quick
proof is obtained by first observing that $\theta$ is singular for
$\tau$ an even integer, and then using modular transformations).
Theta functions have a natural boundary, and are simple examples
of lacunary functions, i.e.\ ``almost all'' their Fourier
coefficients are zero \cite{Apostol:1997}. The partition function
of a compactified time-like boson is thus a formal object from
which it is hard to extract physical meaning. On the other hand,
without compactification, the partition function can easily be
analytically continued. Indeed, replacing the discrete sum by an
integral we have
$$
\int_{-\infty}^{\infty}dx e^{i\pi\tau x^{2}}\ =\ \frac{1}{
\sqrt{-i\tau}}
$$
which can be continued in the lower half plane since it has a
single cut along the negative imaginary axis. \smallskip

We thus restrict to the theory with continuous spectrum of $b$,
and propose the simplest partition function
\begin{equation}
    Z_{k=1}\ =\ \left(|\chi^{R}_{\{0,1/2\}}|^{2}+|\chi^{NS}_{\{0,1/2\}}|^{2}
    +|\chi^{sNS}_{\{0,1/2\}}|^{2}\right)\int_{-\infty}^{\infty}db
    (q\bar{q})^{-b^{2}/2}
    \end{equation}
which we interpret through analytic continuation, up to an
irrelevant phase, as
\begin{eqnarray}
    Z^{phys}_{k=1} & = &
    \left(|\chi^{R}_{\{0,1/2\}}|^{2}+|\chi^{NS}_{\{0,1/2\}}|^{2}
    +|\chi^{sNS}_{\{0,1/2\}}|^{2}\right)\int_{-\infty}^{\infty}d\alpha
    (q\bar{q})^{\alpha^{2}/2} \\[2mm]
 & = & \left(|\chi_{0}|^{2}+|\chi_{1/2}|^{2}\right)^{2}\times \frac{1}{
     \eta\bar{\eta}}\int_{-\infty}^{\infty}d\alpha
     (q\bar{q})^{\alpha^{2}/2}\label{conj}\ \ .
\end{eqnarray}
 This object is obviously modular invariant since, from a direct
 calculation of the integrals,
 \begin{equation}
     \int d\alpha (q\bar{q})^{\alpha^{2}/2}\ =\ \frac{1}{
     \sqrt{\tau\bar{\tau}}}\int d\alpha
     (\tilde{q}\bar{\tilde{q}})^{\alpha^{2}/2}
     \end{equation}
 hence compensating the factors coming from the $\eta$ functions in
 the characters. We note that the spectrum of conformal weights in
 the periodic sector is a continuum starting at $h=1/8$. The
 field with $h=0$ does not appear.
\smallskip

In conclusion, the requirement for our theory to possess a ``physical
partition  function'' has forced us to let $b$ be continuous. Geometrically,
this amounts to a decompactification of the time-like circle. Hence we are
led to consider the universal cover of $\SL21$ so that we can perform an
analytical continuation on the number $b$. One may interpret the
prescription that leads to the physical partition function as an
effective change of the target space along the lines advocated in
\cite{bocquet-2000-578}. Let us also stress that, while our arguments
were based on the $k=1$ theory, it is clear that a similar reasoning
can be carried out for other levels.

The argument leading to $b$ being continuous also seems to exclude 
the smaller theories where part of the complex ${\cal J}$ is dropped, 
at least for general values of $k$. We shall see an exception in the 
case $k=-1/2$ later.

\section{Some selected applications} 

This following section contains some selected applications of our 
general analysis. In the first subsection we shall compare our 
results with studies of the continuum limit of the integrable 
$\sl21\  3 -\bar 3$ spin chain \cite{Essler:2005ag}. The 
agreement we find supports a new interpretation of lattice 
results. The second subsection is devoted to the $k=-1/2$ 
theory which was not included above, but we shall see that 
it shares many of the structures we uncovered throughout 
the last few sections. 

\subsection{The $3-\bar{3}$ super-spin chain revisited.}

In \cite{Essler:2005ag} an integrable $\sl21$ invariant super-spin chain
was studied using both analytical and numerical techniques. Its
Hamiltonian acts on the tensor product $(3\otimes \bar{3})^{\otimes L}$
where $3$ and $\bar 3$ stand for the representations $\{ 1/2\}_{\pm}$ in our
previous terminology. It was argued that in the continuum limit this
chain flows to a $\SL21$ WZNW model at level $k=1$. At the time, the
WZNW model on the supergroup $\SL21$ had not been constructed and it
seems instructive to revisit the issue now on the basis of our
improved understanding of the continuum field theory. We shall see
that the suggested identification with the continuum limit of the
spin chain can be maintained, but some of the lattice results
receive an interesting reinterpretation.
\smallskip

Let us begin by reviewing briefly some results on the spectrum of
the lattice model. In \cite{Essler:2005ag} it was found analytically
that this spectrum exhibits a unique ground state at $h=\bar{h}=0$,
which lies in the single ``true singlet'' of the model, i.e.\ it is
a $\sl21$ invariant state that is not part of a larger indecomposable
representation. This ground state corresponds to an extremely degenerate
solution of the Bethe ansatz equations where all roots collapse to the
origin. Besides the ground state, many excited states were also found.
The lowest lying state above the ground state corresponds to a filled
sea of some (non complex conjugate) string complexes. The rest of the
spectrum is given by excitations obeying the usual pattern of holes
and shifts of the sea. The scaled energies of these excitations over
the ground state were found analytically to be
\begin{eqnarray}\nonumber
     \frac{L}{2\pi v}\Delta E & = &  \frac{1}{ 4} +
    \frac{1}{2}\left(\Delta N_{+}+\Delta
     N_{-}\right)^{2}+\frac{1}{ 8}\left(D_{+}+D_{-}\right)^{2} + \\[2mm]
 &  & \ \ + \  C_N(L)\,  \left(\Delta N_{+}-\Delta N_{-}\right)^{2} +
  C_D(L)\, \left(D_{+}-D_{-}\right)^{2} \label{gaps}\\[3mm]
\mbox{where} & & \ \ C_N(L) \ \rightarrow\  0 \ \ , \ \
C_D(L)\  \rightarrow \ \infty \ \ \mbox{for} \ \
 \ \ \   L\rightarrow\infty\nonumber
\end{eqnarray}
Here, $\Delta N_{\pm},D_{\pm}$ are quantum numbers characterizing the Bethe
ansatz solution. In the continuum limit, the quantity (\ref{gaps}) is
expected to converge to $x=h+\bar{h}$ (all weights in the  $c=0$ theory),
as usual. Formula (\ref{gaps}) indicates an infinite degeneracy of the
level $h=\bar{h}=1/8$ (obtained with $\Delta N_{+}=\Delta N_{-}=D_{+}=
D_{-}=0$ say) in the limit $L\rightarrow\infty$. Numerical studies
confirm this behavior: Indeed, they show that that an infinite number
of levels converges to $h=\bar{h}=1/8$ as $L$ increases. This was
already interpreted in \cite{Essler:2005ag} as indicating the existence of
a continuum of conformal weights starting at $h=\bar{h}=1/8$ in the
thermodynamic limit.
\smallskip

Although an analytical study of the asymptotic corrections to
(\ref{gaps}) seems still out of reach, numerical studies in a
closely related model suggest that the leading contributions
to $C_N$ and $C_L$ can be well fitted by the formulas
\begin{equation} \label{gapsfit}
C_N(L) \ \sim \ \frac{c}{\ln L} + \dots \ \ \ \ , \ \ \ \
C_D(L) \ \sim \ \frac{4}{c}\,  \ln L + \dots
\end{equation}
for large number $2L$ of lattice sites. When these leading terms
are plugged back into the formula (\ref{gaps}) for the spectrum of
the lattice model, we see that the second line resembles very much
the spectrum of a free boson which has been compactified to a
circle with radius square of the order $\ln L$. In other words, if we
assume that eqs.\ (\ref{gapsfit}) are correct, the contribution
from  the ``antisymmetric sector'' (i.e.\ form excitations for
which  $\Delta N_{+}-\Delta N_{-}$ or $D_{+}-D_{-}$ are non
zero) to the partition function can be estimated as
\begin{eqnarray}
     Z_{anti} & = & \frac{1}{ \eta\bar{\eta}}\ \sum_{e,m}\
q^{(e/R+mR/2)^{2}}\ \bar{q}^{(e/R-mR/2)^{2}}\nonumber\\[2mm]
     & = &
     \frac{R}{ \sqrt{2}}\frac{1}{
     \sqrt{\hbox{Im }\tau}\ \eta\bar{\eta}}\ \sum_{m,m'}
   \ \exp\left(-\frac{\pi
     R^{2}|m\tau-m'|^{2}}{ 2\hbox{Im }\tau}\right)
   \nonumber\\[2mm]
     &\approx& \frac{R}{ \sqrt{2}}
     \frac{1}{
     \sqrt{\hbox{Im }\tau}\eta\bar{\eta}}
     \end{eqnarray}
where $\eta(q)$ is Dedekind's eta function, as before. The divergence is
proportional to $R$, i.e.\ to the size of the target space, as expected.
We conclude that in the lattice model, the contribution from the
antisymmetric sector to the partition function multiplies the contribution
from the symmetric sector in eq.\ (\ref{gaps}) by a term of the order of
$\sqrt{\ln L}$. The ground state meanwhile, being a very degenerate Bethe
ansatz solution, does not come with such an extra factor. The generating
function of levels in the periodic sector will therefore have the form
 \begin{equation}
     Z^{R}\ =\ 1+\hbox{cst}\sqrt{\ln L}\ \left[ \frac{1}{
     \sqrt{\hbox{Im }\tau}\ P\bar{P}} (q\bar{q})^{1/8}+\ldots\right]
 \end{equation}
 where the dots represent excitations from the symmetric sector in
 (\ref{gaps}).
\smallskip

The first conclusion we draw is that the contribution of the
continuum completely overrides the one from the discrete state
(as would be the case in any quantum mechanics problem with discrete
states and a continuum with delta function normalizable states), and
that a properly normalized partition function  does not see
the singlet with $h=\bar{h}=0$. The resulting object is in good
agreement with our conjectured partition function (\ref{conj}).%
\smallskip

Reference \cite{Essler:2005ag} contained various failed attempts to
build a conformal field theory containing both the continuum of
representations $\{b,j=1/2\}$ and a single identity field associated
with the representation $\{0\}$. Given our new insight into the
continuum model, the problems to incorporate the singlet state
may not come as a complete surprise. Although the 
free field construction at $k=1$ suggests the possibility of smaller 
theories, the study of modular invariants (as well as of four point 
functions, as we mentioned above) seems to preclude the appearance 
of the  singlet representation on its own - i.e.\ without being 
part of a big indecomposable with vanishing super-dimension. In 
addition all the states we found in the continuum approach were 
non normalizable. Both observations lead us to speculate that
eq.\ (\ref{conj}) represents the full operator content of the
continuum limit, and that there is no discrete state associated
with a true singlet. Put differently, the new investigation
suggests that the true singlet observed on the lattice is an
artifact of the regularization and does not belong to the
continuum limit. Our new interpretation of the lattice results
receives additional support from the very singular nature of
the Bethe ansatz solution that corresponds to the singlet
state. It would be interesting to check further the decoupling of the 
true singlet by studying the scaling behavior of
matrix elements of lattice regularized 
current algebra generators. 
\smallskip

There is one more potential objection one might raise. Note that
in our continuum theory fermionic and bosonic states are perfectly
paired so that the Witten index of the $\SL21$ WZNW model is
guaranteed to vanish. Meanwhile, for our lattice spin chain on
the space $(3\otimes \bar{3})^{\otimes L}$ one finds an excess
by one for the number of bosonic states over the number of
fermionic ones. Hence, the Witten index is non-zero on the
lattice and one would naively expect the same to be true for
the continuum limit, in conflict with what we have proposed
above. In order to resolve this issue, we suggest
that there exist different spin chains which give rise to  the
same continuum limit while possessing an excess of fermionic
states over bosonic. More concretely, while we do not understand
the whole structure yet, we have found \footnote{We thank F. Essler 
for kindly exploring this question numerically.}
that the ground states of
integrable chains of the type $(3\otimes \bar{3})^{\otimes L}
\otimes 3$ and $\bar{3}\otimes (3\otimes \bar{3})^{\otimes L}$
scale to conformal weight $h=0$ as well (in fact the ground
state energy is given exactly by $E_{0}=-\hbox{length}\times
~ e_{0}$ where $e_{0}$ has no finite size correction and is the
same for all chains), but this time they come in the
representation $3$ (resp.\ $\bar{3}$). Once we sum over the
various lattice models, the balance between bosonic and
fermionic states may be restored even before taking the
continuum limit.

\subsection{The WZNW model at $k=-1/2$}

Our investigation above was restricted to integer level $k$. But
as we have mentioned before, these are some fractional values of
$k$, in particular $k=-1/2$, which play an important role for
applications. While we are not prepared to give a systematic
account on fractional level theories, we would like to discuss
briefly a model with $k=-1/2$. Our analysis will lead to the
remarkable conclusion that the basic structure of this model is
essentially the same as for integer $k$, only that  there exist 
several components within the atypical sector, each of them 
being modeled after ${\cal J}$. 
\medskip

In this case $k=-1/2$, the relevant representation theory of the
$\sl21$ current algebra is particularly simple. In fact, all
relevant representations can be obtained from the vacuum sector
$\{0\}^\wedge$ through application of spectral flow symmetries. It
is not difficult to show that at $k=-1/2$ the automorphism
$\gamma^2$ is inner, i.e.\ $\gamma^2 \sim id$. This means that
application of $\gamma^2$ does not lead to any new
representations. The remaining nontrivial automorphisms are of the
form $\gamma^n_+ \gamma^\sigma$ with $n \in \mathbb{Z}$ and
$\sigma = 0,1$. We shall denote the corresponding irreducible
representations of the $\sl21$ current algebra by
$$  \{(n,\sigma)\} \ \cong \ \gamma^n_+\, \gamma^\sigma\{0\}^\wedge
\ \ \ .
$$
By construction, this set closes under fusion. In fact, the fusion
product simply amounts to a composition of the associated
automorphism.
\smallskip

With the exception of the sectors labeled by $n=0,\pm 1$, the
representations $\{(n,\sigma)\}$ do not contain a highest or
lowest weight. The representation $\{(0,0)\}$ is to be identified
with the vacuum representation. $\{(0,1)\} = \{0,1/2\}^\wedge$ is
the only other admissible representation at $k=-1/2$. It is
generated from the 4-dimensional typical multiplet $\{0,1/2\}$ of
ground states with conformal weight $h = 1/2$. In addition, there
are four more highest/lowest weight weight representations which
are erected over the atypical discrete series representations
$\{(-1,\sigma_\pm)\} = \{(-,-1/4)\}^\wedge_\pm$ and
$\{(1,\sigma_\pm\} = \{(+,-1/4)\}^\wedge_\pm$ corresponding to a
negative spin $j=-1/4$. The choice of the sign in the first
argument of the bracket determines on whether the representation
is highest $(-)$ or lowest $(+)$ weight. The subscript, on the
other hand corresponds to the two different choices of the
parameter $b$ that make these representations atypical. All four
representations possess ground states of conformal weight $h=0$.
In all other representations $\{(n,\sigma)\}$ with $|n| \geq 2$, 
the conformal weight is unbounded from below.
\smallskip

Since we can generate every representations from $\{ 0\}^\wedge$,
is suffices to display the character of the vacuum representation, 
$$ \chi_{\{0\}}(q,z,\xi) \ = \ \frac12 \left[
\frac{\vartheta_3(q,\xi^{1/2})}{\vartheta_4(q,z^{1/2})} +
\frac{\vartheta_4(q,\xi^{1/2})}{\vartheta_3(q,z^{1/2})} \right] \
\ . $$ We shall explain the origin of this formula in a moment.
Characters of all the other representations are obtained from the
vacuum character $\chi_{\{0\}}$ through
$$ \chi_{\{(n,\sigma)\}}(q,z,\xi)\ = \ \gamma^n_+\, \gamma^\sigma\,
  \chi_{\{0\}}(q,z,\xi) \ = \ \xi^{-\frac{n}{4} -\frac{\sigma}{2}}
  \, z^{-\frac{n}{4}}
  q^{\frac{n+1}{2}\sigma}\,
  \chi_{\{0\}}(q,q^{-n-2\sigma}\xi,q^{nz}z)\ \ .
$$
To derive the above character formula and for the subsequent
discussion we note that the $\sl21$ current algebra at level
$k=-1/2$ possesses a free field representation which employs the
same free fields as in the case of the $k=1$ theory, i.e.\ two
free bosonic fields $\phi$ and $\phi'$ with space-like and
time-like signature, respectively, and a pair of symplectic
fermions $\eta_1,\eta_2$. The bosonic $\sl21$ currents read
\begin{eqnarray}
    E^{+}(z) \ =\ \frac12 \, e^{-2i\varphi'(z)}\, \partial^{2}\eta_{1}(z)
\, \partial\eta_{1}(z) \ \ \ & , & \ \ \ \ H(z) \ = \
\frac{i}{2}\,
   \partial\phi'(z) \\[2mm]
   E^{-}(z)\ =\ \frac12\, e^{2i\varphi'(z)}\, \partial^{2}\eta_{2}(z)\,
   \partial\eta_{2}(z) \ \ \ & , & \ \ \ \ B(z) \ = \ \frac{i}{2}\,
    \partial \phi(z)\ \ \ .
\end{eqnarray}
Note that, unlike in the case of $k=1$, the bosonic currents
involve the symplectic fermions and the time-like free boson. For
the fermionic currents one finds
\begin{eqnarray}
    V^{-}(z)\ =\ \frac{1}{\sqrt{2}} \,
    e^{-i(\varphi(z)-\varphi'(z))}\, \partial\eta_{2}(z) \ \ \ \ & , &
    \ \ \ \
    V^{+}(z)\ =\ \frac{1}{\sqrt{2}}\,
       e^{-i(\varphi(z)+\varphi'(z))}\, \partial\eta_{1}(z) \nonumber\\[2mm]
       W^{-}(z) \ =\ \frac{1}{\sqrt{2}}\,
      e^{i(\varphi(z)+\varphi'(z))}\, \partial\eta_{2}(z)  \ \ \ \ & , &
    \ \ \ \   W^{+}(z)\ =\ \frac{1}{\sqrt{2}}\,
      e^{i(\varphi(z)-\varphi'(z))}\, \partial\eta_{1}(z) \ \ . \nonumber
\end{eqnarray}
As in the case of the $k=1$ theory, the free field construction
determines a consistent model with a $\sl21$ current algebra
symmetry. If we do not include the symplectic fermions (note that
once more the currents only involve derivatives), but only their
derivatives then the state space reads
$$ {\cal H}_{k=-1/2} \ = \ \bigoplus_{n,\sigma} \
   \{(n,\sigma)\}^\wedge \times \{(-n,\sigma)\}^\wedge\ \ .
   $$
Since the spectral flow automorphisms $\gamma_\pm$ and $\gamma$
correspond to multiplication with the fields
$$ \gamma_\pm \ \leftrightarrow \ e^{-\frac{i}{2}(\phi'\pm\phi)}
\ \ \ , \ \ \ \gamma \ \leftrightarrow \ e^{- i \phi }  \ \ $$
is is fairly easy to write down at least one field in each
sector of the model,
$$ \{(n,\sigma)\}^\wedge \times \{(-n,\sigma)\}^\wedge
 \ \ \ \mbox{contains} \ \ \  e^{-i\frac{n}{2}\phi' -
   i\frac{n+2\sigma}{2}\phi} \ \ . $$
The space ${\cal H}_{k=-1/2}$ contains R sector representations
only, but it is certainly possible to include the NS sector by
adding the image under the spectral flow $\vartheta$. Since this
works just in the same way as above, we shall not repeat the
discussion here.
\smallskip

Even though all the representations we are working with are
atypical, the state space decomposes into irreducible building
blocks. This is quite different from the structure of the atypical
sectors ${\cal J}^\wedge$ we described above. On the other hand,
is is very similar to one of the consistent theories with $k=1$
that we described at the end of section 4.2. In the $k=1$ theory,
the singular vectors of the indecomposable block ${\cal J}^\wedge$
were decoupled by restricting to a theory that contained only
derivatives of the fermionic fields. Conversely, the experience
from $k=1$ suggest that in the $k=-1/2$ case we may be able to
construct a theory with a more complicated atypical sector by
including one or both of the symplectic fermion fields
$\eta_1$ and $\eta_2$ \cite{Rasmussen}.
\smallskip

We claim that in case we include both fermionic zero modes we end
up with an atypical structure that decomposes into four different
blocks, each of them being built in the same way as our sector
${\cal J}^\wedge$. We shall present the analysis only for the
block that contains the vacuum sector $\{ 0 \}^\wedge$. The other
three sectors are obtained by acting with $\gamma, \gamma_+$ and
$\gamma_+ \circ \gamma$. Let us start our discussion with the
field $\eta_1 \eta_2$. Any action with $V^\pm, W^\pm$ and $E^\pm$
will remove one of the two fermionic zero modes and hence $\eta_1
\eta_2$ sits at the top of a sector $\{ 0 \}^\wedge$. The
action of $V^+,W^+$ and $E^+$ takes us from here into a set of
fields which all contain a factor $\eta_1$. These fields can be
shown to belong to a sector that is isomorphic to $\{(2,0)\} \cong
\gamma_+^2 \{0\}^\wedge$. Further application of  $V^-,W^-$ and
$E^-$ bring us to a set of fields that contain only derivatives of
fermions. These form a subrepresentation $\{0\}^\wedge$ at the
bottom of our atypical representation. A similar analysis applies
if we act with $V^-,W^-$ and $E^-$ first. This time, we descend to
$\{0\}^\wedge$ via the sector $\{(-2,0)\} \cong
\gamma_+^{-2}\{0\}^\wedge$. Continuing along this line of
thoughts, one can see that the sectors $\{(2n,0)\}, n \in
\mathbb{Z},$ form the composition series for an indecomposable
representation ${\cal J}^\wedge$ with $\{(2n,0)\}$ in place of
$\{n/2\}$. The state space of the maximal theory therefore
decomposes into four indecomposable blocks. Once more, there are
two intermediate theories, each of which has four saw-blade shaped
atypical sectors. They are obtained if we omit either $\eta_1$ or
$\eta_2$ (but not their derivatives, of course) from the above
maximal theory.
\smallskip

Even though we are not prepared to analyze WZNW models for generic
fractional levels, it is remarkable that the structure we have
first uncovered in our minisuperspace limit, re-appears even for
$k = -1/2$. It seems very likely that the same is true for a
generic choice of the level.

\section{Conclusions and Outlook}

An obvious conclusion of our study is that WZNW models on supergroups
are interesting examples of logarithmic CFTs, much richer than it has
been anticipated in earlier works. Gurarie (see \cite{Gurarie:2004ce}
and references therein), for instance, argued that super WZNW models
with $c=0$ could be considered as made of two ``decoupled'' component
theories with opposite values of the central charge, an observation
justified in part by the fact that in the GL(1$|$1) WZNW model, the
stress energy tensor belongs naturally to a four dimensional GL(1$|$1)
multiplet in which $L_{0}$ is diagonalizable, and hence $T$ has no
``non trivial logarithmic partner''. The SU(2$|$1) WZNW clearly does
not obey any such decoupling. In fact, restricting to the right
moving current algebra as in \cite{Gurarie:2004ce} we see that the
identity field belongs to a projective representation of the zero
mode algebra on which the Casimir - and hence $L_{0}$  - is not
diagonalizable. Applying $L_{-2}$ to this representation produces
a  Virasoro Jordan cell at level $h=2$ and a non trivial
logarithmic partner of the stress energy tensor. This can be seen
quite explicitely in the case $k=1$ where, within the free field
representation (and similarly to the case of symplectic fermions), 
the field
$$ t(z) \ := \ :\eta_{1}(z)\eta_{2}(z) T(z): $$
is a logarithmic partner of
$$T\ :=\  :\partial\eta_{1}(z)\partial\eta_{2}(z): -
   \frac{1}{ 2}:(\partial\phi(z))^{2}:
+\frac{1}{ 2}:(\partial\phi'(z))^{2}: \ \ .
$$
Note that the whole structure of indecomposables is in fact much
more complicated than envisioned in \cite{Gurarie:2004ce} when
the interplay of left and right current algebras is taken into
account.
\smallskip 

Even though the structure of the state space is rather difficult 
when analyzed with respect to the combined left and right action, 
it is surprisingly simple once we restrict to either the left or 
the right action alone. Note that the Lie superalgebra $\sl21$ 
has a large number of indecomposables (see e.g. \cite{Gotz:2005jz}) 
from which only a very distinguished sub-class does actually occur 
within the state space of our model. In fact, we have seen 
above that all states (both in the minisuperspace theory and the 
full field theory) transform according to the so-called projective 
representations of $\sl21$, i.e.\ either in typicals and projective 
covers of atypicals. This is not to say, however, that other 
representations of $\sl21$ have no relevance for sigma models on 
supergroups. In addition to the left and right regular representation  
there is yet one more important symmetry that arises from the adjoint 
action of $\sl21$ on the state space. With respect to the latter, 
states can transform in other indecomposables. The underlying 
mathematical structure turns out to be quite intriguing and will 
be described elsewhere. Since the adjoint action is left unbroken 
by maximally symmetric boundary conditions, the resulting 
decomposition should have applications, in particular to the 
study of boundary conditions for sigma models on supergroups. 
\smallskip 

As a final comment let us point out one generic feature we have 
encountered in both $\GL11$ and $\SL21$, namely that the contribution 
of the indecomposable sector ${\cal J}$ simply 
makes up for the subtractions in the atypical sectors of the theory, 
so that  the partition function sees only contributions from   Kac 
modules, and has a simple factorized form. This behavior is sufficient 
for a modular invariant partition function but it is not necessary.   
The potential existence of different versions of the theory where only 
parts of the complex ${\cal J}$ appear, requires more study. We note 
that some hints in this direction are provided by the study of four 
point functions. In the case $k=1$, the four point function of the 
fields in the $\{0,1/2\}$ representation has been studied in detail. 
It turns out that the KZ equations factorize, and that it is possible 
to decouple one conformal block. Two blocks remain, leading to 
logarithmic dependence, and indicating that the identity field 
remains part of an indecomposable representation. This suggests 
that the smallest theory, where the complex ${\cal J}$ is reduced 
to an infinite sum of irreducibles, cannot appear in the $\sl21$ 
WZNW model. In the case $k=-1/2$ meanwhile it is possible to 
decouple two conformal blocks, leaving only the identity field, 
and indicating that the smallest theory does make sense this time 
- a feature consistent with the free field representation and the 
modular invariant. The $\sl21$ WZNW model at fractional level and 
the explicit construction of consistent theories with a truncated
atypical sector certainly deserve a more systematic investigation. 
\bigskip \bigskip 

\noindent
{\bf Acknowledgments:} We thank Fabian Essler, Gerhard G\"otz, Thomas Quella
and Anne Taormina for interesting conversations. V.S. would like 
to thank the SPhT for the warm hospitality during several stays. 
This work was partially supported by the EU Research Training 
Network grants ``Euclid'', contract number HPRN-CT-2002-00325
and ``ForcesUniverse'', contract number MRTN-CT-2004-005104.

\section{Appendix A: The right regular representation}

In this appendix we would like to prove the decomposition formula
for the right regular representation. We shall use the same
notations that were introduced in section 2.2. In order to analyze
the decomposition of the space of functions under the right
regular action of $\sl21$, we shall first study its restriction to
the Lie sub-superalgebra $\gl11$. More precisely, we shall make
use of the following embedding
$$ \e(\psi^+) \ = \ F^+ \ \ , \ \ \e(\psi^-) \ = \ \bar F^- \ \ , \ \
\e(E) \ = \ B-H \ \ , \ \ \e(N) \ = \ B+H\ \ .
$$
The main technical Lemma of this section implies that under the
action of $\gl11$, the space ${\cal H}$ of functions on the
supergroup $\SL21$ decomposes into projectives only.
\medskip

\noindent {\bf Lemma: } {\it Under the action of $R_X \equiv
R_{\e(X)}$ of the generators $X \in$ $\gl11$, the space ${\cal H}$
of functions of $\SL21$ decomposes according to
$$  {\cal H} \ \cong \ \bigoplus_j \bigoplus_{b=-j}^j
  {\cal P}(2b+1) \oplus 2 \cdot {\cal P}(2b) \oplus {\cal
  P}(2b-1) \oplus \ {\cal T} \ .
$$
Here, ${\cal T}$ is a direct sum of typical $\gl11$ representations
and $\P(a)$ denotes the projective cover of the atypical
irreducible $\langle a\rangle$.}
\medskip

\noindent Before we prove this statement, let us formulate two
consequences for the right regular representation of $\sl21$. To
begin with, let us recall from \cite{Gotz:2005jz} that an
$\sl21$ representation $\pi$ descends on a projective representation
of the embedded $\gl11$ algebra if and only if $\pi$ is projective.
Our lemma claims that the $\gl11$ action on {\cal H} contains only
projectives. Hence, the same must be true for the right regular
action of $\sl21$.
\bigskip

\noindent {\sc Proof of Lemma:} For the proof it will be useful to
introduce the following odd functions
$$ \bar\theta_\pm \ := \ e^{\mp i z/2} D^{1/2}_{(\pm 1/2)\nu}(g^{-1})
    \ \bar \eta_\nu \ \ \ , \ \ \ \tilde \eta_- \ = \ e^{iz
    }
    \eta_- \ \ .
$$
It is not difficult to see that the space ${\cal H}$ is spanned by
functions of the form
$$ F^{n,j}_{ab} \Lambda \ = \ e^{inz} \ D^j_{ab}(g) \
    \Lambda(\tilde\eta_-,\eta_+,\bar\theta_\pm)
\ \ , $$ where $\Lambda(\tilde\eta_-,\eta_+,\bar\theta_\pm)$ is an
arbitrary element in the algebra generated by the arguments. It is
very easy to describe explicitly the space of functions which are
organized in atypicals of $\gl11$. The latter is characterized by
the vanishing of $R_E$,
\begin{equation}
 {\cal A}_R \ = \ \{  \Phi \in {\cal H} \ | \
 \left(i\partial_z + R^0_h + \eta_- \partial_-\right)\Phi \ = \ 0
 \} \ \ .
\end{equation}
We can easily solve the equation for $\Phi$ and describe the space
${\cal A}_R$ explicitly. In fact, it is spanned by the functions
$$ F^{b,j}_{ab} \Lambda \ = \ e^{ibz} D^j_{ab}(g) \
\Lambda(\tilde\eta_-,\eta_+,\bar\theta_\pm) \ \ . $$ On the
subspace ${\cal A}^R$ the other generators of $\gl11$ simplify to
\begin{eqnarray}
R_N \Phi & = & \left( -2i\partial_z  -\eta_- \partial_-
           -\eta_+ \partial_+ \right) \Phi  \\[2mm]
R_{\Psi^+} \Phi \ = \ -i \partial_+ \Phi\ \  &,& \ \ R_{\Psi^-}
\Phi \ = \ i e^{-iz/2} \,
                    D^{1/2}_{\nu (- 1/2)} (g)\ \bar \partial_{-\nu}
                    -  i \eta_- R^0_{E^-} \ .
\end{eqnarray} for all $\Phi \in {\cal A}_R$. The representation
of $\gl11$ can be restricted to the space ${\cal A}'_R$ of all
elements $\Phi \in {\cal A}_R$ such that $\eta_-\partial_- \Phi =
\Phi$. A short look on the action of the $\gl11$ generators reveals
that
$$ {\cal A}'_R \ = \ \bigoplus_j \bigoplus_{b=-j}^j
  {\cal P}(2b+1) \oplus {\cal P}(2b) \ \ . $$
Similarly, we see that
$$ {\cal A}_R/{\cal A}'_R \ = \ \bigoplus_j \bigoplus_{b=-j}^j
  {\cal P}(2b) \oplus {\cal P}(2b-1)\ \ .
$$
Since all representations are projective we conclude that
$$ {\cal A}_R \ = \ \bigoplus_j \bigoplus_{b=-j}^j
  {\cal P}(2b+1) \oplus 2 \cdot {\cal P}(2b) \oplus {\cal
  P}(2b-1)\ \ .
$$
This concludes the proof of our Lemma. \bigskip

\def\cprime{$'$} \def\cprime{$'$}
\providecommand{\href}[2]{#2}\begingroup\raggedright

\endgroup

\end{document}